\documentclass[12 pt]{article}
\usepackage[a4paper, total={7.5in, 8in}]{geometry}
\usepackage{graphicx} 
\usepackage{tikz}
\usepackage{subcaption} 
\usepackage{amsmath}
\usepackage{amssymb}
\usepackage{float}
\usepackage{multicol}
\usepackage{multirow}

\usepackage[style=authoryear, 
            sorting=nyt, 
            maxcitenames=2, 
            maxbibnames=99, 
            uniquename=false]{biblatex}

\newtheorem{Theorem}{Theorem}
\newtheorem{lemma}{Lemma}
\newtheorem{Proposition}{Proposition}
\newtheorem{Corollary}{Corollary}

\newcommand{\E}{{\mathbb{E}}}

\title{Adaptive Proximal Causal Inference with Some Invalid Proxies}
\author{Prabrisha Rakshit, Xu Shi, Eric Tchetgen Tchetgen}
\date{July 2023}

\begin{document}

\maketitle

\begin{abstract}
    Proximal causal inference (PCI) is a recently proposed framework to identify and estimate the causal effect of an exposure on a given outcome, in the presence of hidden confounders for which proxies are available. Specifically, PCI relies on having observed two valid types of proxies; a \emph{treatment-inducing confounding proxy} related to the outcome only to the extent that it is associated with an unmeasured confounder conditional on the primary treatment and measured covariates, and an \emph{outcome-inducing confounding proxy} related to the treatment only through its association with an unmeasured confounder conditional on measured covariates. Therefore, valid proxies must satisfy stringent exclusion restrictions; mainly, a treatment-inducing confounding proxy must not cause the outcome, while an outcome-inducing confounding proxy must not be caused by the treatment. In order to improve the prospects for identification and possibly the efficiency of the approach, multiple proxies will often be used, raising concerns about bias due to a possible violation of the required exclusion restrictions. To address this concern, we introduce necessary and sufficient conditions for identifying causal effects in the presence of many confounding proxies, some of which may be invalid. Specifically, under a canonical proximal linear structural equations model, we propose a LASSO-based median estimator of the causal effect of primary interest, which simultaneously selects valid proxies and estimates the causal effect with corresponding theoretical performance guarantees. Despite its strengths, the LASSO-based approach can under certain conditions lead to inconsistent treatment proxy selection. To overcome this limitation, we introduce an adaptive LASSO-based proximal estimator, which incorporates adaptive weights to differentially penalize separate treatment proxy coefficients with respect to the $\ell_1$ penalty. We formally establish that the adaptive estimator is $\sqrt{n}$-consistent for the causal effect, and when a valid outcome-confounding proxy is available, we construct corresponding asymptotically valid confidence intervals for the causal effect. We also extend the approach to the many outcome-confounding proxies setting, some of which may be invalid. All theoretical results are supported by extensive simulation studies. We apply the proposed methods to assess the impact of right heart catheterization on 30-day survival outcomes for critically ill ICU patients, utilizing data from the SUPPORT study. 
\end{abstract}

\section{Introduction}
\label{sec: intro}

Causal inference from observational data has historically primarily relied on the key untestable assumption that one has accurately measured all relevant confounding variables, such that units are exchangeable across treatment arms, conditional on measured covariates, so-called conditional exchangeability. However, even in well-designed observational studies where all important sources of confounding might be known, it may be unrealistic to assume that relevant confounders can be measured perfectly. In such settings, measured variables may at best be viewed as confounding proxies, which, although are associated with hidden confounding factors, do not suffice to ensure conditional exchangeability.  Proximal causal inference was recently proposed 
\parencite{introeric, ericbio} as a formal framework to identify and estimate causal effects in the presence of hidden confounders for which valid proxies are available. The approach technically relies on having two types of proxies of hidden confounders available in the observed sample: \emph{treatment-inducing confounding proxies (TCP),} and \emph{outcome-inducing confounding proxies (OCP)}, which meet certain stringent exclusion restriction conditions.  Basically, 
a valid TCP must not have a direct effect on the primary outcome, and therefore must be associated with the latter only to the extent that it is associated with an unmeasured confounder conditional on the treatment and measured confounders; while an OCP must not directly be causally impacted by the primary treatment, and therefore be associated with the latter only to the extent that it is associated with a hidden confounder to the outcome solely through an unmeasured common cause for which the variable acts as a proxy, conditional on measured confounders. Furthermore, TCP and OCP variables must be conditionally independent given the primary treatment of interest, measured, and unmeasured confounders. 

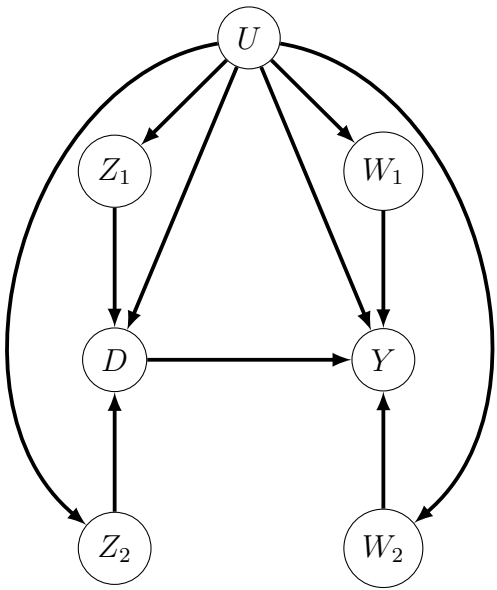
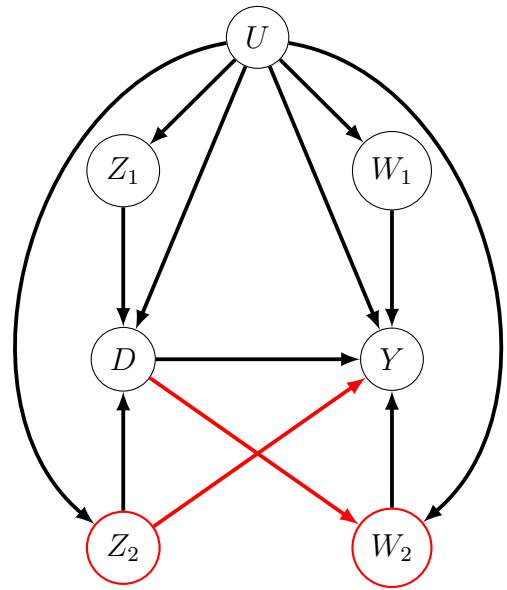
\begin{figure}[ht]
    \centering
    \begin{subfigure}[b]{0.45\textwidth}
        \centering
        \begin{tikzpicture}[node distance={25mm}, main/.style = {draw, circle}, rednode/.style = {draw=red, circle, thick}] 
            \node[main] (1) {$U$};
            \node[main] (2) [below left of = 1] {$Z_1$};
            \node[main] (3) [below right of = 1] {$W_1$};
            \node[main] (4) [below of = 2] {$D$};
            \node[main] (5) [below of = 3] {$Y$};
            \node[main] (6) [below of = 4] {$Z_2$};
            \node[main] (7) [below of = 5] {$W_2$};

            \draw[->, >=latex, line width=0.5mm] (1) -- (2);
            \draw[->, >=latex, line width=0.5mm] (1) -- (3);
            \draw[->, >=latex, line width=0.5mm] (2) -- (4);
            \draw[->, >=latex, line width=0.5mm] (3) -- (5);
            \draw[->, >=latex, line width=0.5mm] (4) -- (5);
            \draw[->, >=latex, line width=0.5mm] (1) -- (4);
            \draw[->, >=latex, line width=0.5mm] (1) -- (5);
            \draw[->, >=latex, line width=0.5mm] (6) -- (4);
            \draw[->, >=latex, line width=0.5mm] (7) -- (5);
            \draw[->, >=latex, line width=0.5mm, bend right=65] (1) to (6);
            \draw[->, >=latex, line width=0.5mm, bend left=65] (1) to (7);
        \end{tikzpicture}
        \caption{$Z_1, Z2$ and $W_1, W_2$ are valid TCP and OCP respectively.}
        \label{fig: valid proxies}
    \end{subfigure}
    \hfill
    \begin{subfigure}[b]{0.45\textwidth}
        \centering
        \begin{tikzpicture}[node distance={25mm}, main/.style = {draw, circle}, rednode/.style = {draw=red, circle, thick}] 
            \node[main] (1) {$U$};
            \node[main] (2) [below left of = 1] {$Z_1$};
            \node[main] (3) [below right of = 1] {$W_1$};
            \node[main] (4) [below of = 2] {$D$};
            \node[main] (5) [below of = 3] {$Y$};
            \node[rednode] (6) [below of = 4] {$Z_2$};
            \node[rednode] (7) [below of = 5] {$W_2$};
            \draw[->, >=latex, line width=0.5mm] (1) -- (2);
            \draw[->, >=latex, line width=0.5mm] (1) -- (3);
            \draw[->, >=latex, line width=0.5mm] (2) -- (4);
            \draw[->, >=latex, line width=0.5mm] (3) -- (5);
            \draw[->, >=latex, line width=0.5mm] (4) -- (5);
            \draw[->, >=latex, line width=0.5mm] (1) -- (4);
            \draw[->, >=latex, line width=0.5mm] (1) -- (5);
            \draw[->, >=latex, line width=0.5mm] (6) -- (4);
            \draw[->, >=latex, line width=0.5mm] (7) -- (5);
            \draw[->, >=latex, line width=0.5mm, red] (6) -- (5);
            \draw[->, >=latex, line width=0.5mm, red] (4) -- (7);
            \draw[->, >=latex, line width=0.5mm, bend right=65] (1) to (6);
            \draw[->, >=latex, line width=0.5mm, bend left=65] (1) to (7);
        \end{tikzpicture}
        \caption{$Z_2$ and $W_2$ are invalid proxies, indicated by the presence of the red arrows in the DAG.}
        \label{fig: invalid proxies}
    \end{subfigure}
    \caption{Comparison of two DAGs with different proxy validity.}
\end{figure}

Statistical methods for proximal causal inference with valid proxies are now well developed, in both parametric and semiparametric/nonparametric settings \parencite{semiparam, ghassami, deaner, jiewen}. 
%
To ground ideas, we briefly describe a standard PCI approach with valid confounding proxies, under a pair of linear structural equations for the primary outcome $Y$ and OCP $W$, in terms of the primary treatment $D$, TCP $Z$ and hidden confounder $U$:
\begin{equation}
\begin{aligned}
    \mathbb{E}(Y \mid D, Z, U) &= \beta D + \beta_u U, \\
    \mathbb{E}(W \mid D, Z, U) &= \eta_u U.
\end{aligned}    
\end{equation}
where as a valid TCP $Z$ appears in the conditioning event of both equations but not on their right-hand side; while as a valid OCP, the conditional mean of $W$ also does not depend on $A$ conditional on $U$. Assuming $W$ is $U$-relevant, that is $\eta_u \neq 0$, we have that:
\begin{equation}
\mathbb{E}(Y \mid D, Z) = \beta D + \frac{\beta_u}{\eta_u} \mathbb{E}(W \mid D, Z).    
\end{equation}
Thus, given a consistent estimator $\widehat{W}$ of $\mathbb{E}(W \mid D, Z)$, a least-squares regression of $Y$ on $D$ and $\widehat{W}$ should recover a consistent estimator of $\beta$, the causal parameter of interest. This so-called ``proximal two-stage least squares" (P2SLS) method is analogous to the 2SLS approach used in instrumental variable (IV) settings, allowing existing instrumental variable software to be re-purposed for P2SLS by treating $W$ as the endogenous variable, $Z$ as the instrument, and $D$ as a covariate. 
Figure \ref{fig: valid proxies} illustrates the conditional independencies that valid TCP and OCP variables must satisfy to formally justify PCI. 

In an effort to improve the prospects for identification and potentially improve efficiency, multiple TCPs and OCPs are often considered in practical applications of proximal causal inference, raising concerns about bias due to a possible violation of the required exclusion restrictions. Specifically, existing PCI methods with multiple proxies require all proxies to be valid and may be severely biased even if a single candidate proxy does not fulfill the required exclusion identifying conditions, that is, say, if a TCP causally impacts the primary outcome of interest, or if the primary treatment has a direct causal effect on an OCP. Figure \ref{fig: invalid proxies} illustrates potential violations of proxy conditions that would in principle invalidate the above P2SLS approach.   To address this concern, in this paper, we derive the necessary and sufficient conditions for identifying causal effects in the presence of many confounding proxies, some of which may not be valid.  Our main contributions are outlined in the following Section. 

        
        


\section{Main Contributions and Related Works}

Proximal causal inference has recently emerged as a formal framework to leverage proxy variables to account for hidden confounders for which they are relevant.  Originally proposed to address unmeasured confounding in a point exposure setting 
(\cite{ericbio,introeric}), the framework has since been extended in multiple directions, including mediation analysis subject to unmeasured confounding \parencite{dukemed}, 
mediation analysis with a hidden mediator \parencite{ghassamimed}, 
time-varying treatment subject to time-varying confounding \parencite{introeric, yinglong}, 
and causal inference on interconnected units subject to hidden homophily \parencite{egami} .
Additional related methodological developments on proximal causal inference include  \cite{ridge},  
\cite{shi2020multiply}, \cite{semiparam}, \cite{ghassami}, \cite{kallusminimax} and \cite{deaner}.

\vspace{2 mm}
\noindent As mentioned previously, in practice, one may wish to leverage several available candidate proxies to improve efficiency and the prospects for identification; however, such a strategy may open one  up to potential bias due to one or more invalid candidate proxies violating the required exclusion restrictions.  In this work, we depart from existing PCI literature and consider formal conditions for obtaining valid causal inferences about a treatment effect subject to confounding by hidden factors for which several candidate proxies may be available, some of which unbeknownst to the analyst are invalid. Despite the fast-evolving proxy literature, to the best of our knowledge, existing works have exclusively relied on the crucial assumption that available candidate proxy variables perfectly satisfy the required exclusion restrictions. A notable exception to the standard assumptions is the DANCE algorithm proposed by \cite{DANCE}, which data-adaptively identifies a special class of proxy variables. While useful, DANCE is limited to proxies satisfying a specific structural condition. Another recent work by \cite{invTCP} introduces a semiparametric framework that accommodates invalid TCPs through a new class of bridge functions and develops estimators for the average treatment effect. However, their focus remains on potentially invalid TCPs alone.

\vspace{2 mm}
\noindent In contrast, our work considers both invalid TCPs and invalid OCPs under a  flexible structural framework. 
We begin by assuming the availability of a valid OCP and develop a method to identify valid TCPs among many candidates, some of which may be invalid, without prior knowledge of their validity. We then relax this assumption and consider the more challenging setting in which OCPs themselves may also be invalid. Specifically, focusing on a many-proxies structural linear equations model, we adapt and extend recently developed methods by \cite{kang2016, windmeijer2019} for robust causal inference with many instrumental variables, some of which are invalid,  to the current task of obtaining robust inferences about a treatment effect with many proxies, some of which are invalid. Although \cite{kang2016, windmeijer2019} introduced the use of LASSO to select valid instruments, in contrast, our proposed methodology uses LASSO to select valid TCPs, which presents several new challenges that require novel technical arguments due to key structural differences between valid instruments compared with valid proxies, and the distinct role performed by valid TCP and OCP variables in proximal inference. Mainly, a valid instrument must be independent of all unmeasured confounders, while valid proxies should ideally be directly related to hidden confounders; furthermore TCP and OCP must be independent conditional on the treatment and hidden confounders, an assumption that does not figure into instrumental variable analysis.  In light of these differences with prior literature, our main contributions are as follows:

\begin{enumerate}
    \item {\bf PCI with unknown invalid TCPs:} 
    We address the challenge of estimating the causal effect of a treatment on an outcome of interest in the presence of unmeasured confounding, leveraging a set of possibly invalid candidate proxies of hidden confounders, without apriori knowing which of the proxies are valid. Specifically, we consider increasingly challenging conditions, starting with a setting where a single valid OCP is available, alongside a pool of TCPs, some of which may be invalid. We propose a LASSO-based median estimator of the causal effect of primary interest, which simultaneously selects valid proxies from a pool of candidate proxies, and we obtain an estimator of the causal effect of primary interest while achieving appealing theoretical performance guarantees. Despite exhibiting certain favorable robustness properties, the LASSO-based approach can under certain conditions lead to inconsistent treatment proxy selection. To overcome this limitation, we introduce an adaptive LASSO-based proximal estimator, which incorporates adaptive weights that differentially penalize separate treatment proxy coefficients with respect to the $\ell_1$ penalty. We formally establish that the adaptive estimator is $\sqrt{n}$-consistent for the causal effect of primary interest, with Gaussian asymptotic distribution which matches that of an oracle with apriori knowledge of the set of valid TCPs. 

    \item {\bf PCI with both invalid TCPs and OCPs:} The paper next considers a setting in which both candidate TCPs and OCPs may be invalid. We propose a median estimator which carefully incorporates the adaptive LASSO estimator described above. We provide conditions under which the proposed estimator is $\sqrt{n}$-consistent. To address inference, we adopt a subsampling-based approach to construct valid confidence intervals for the causal effect in view,  
    therefore further enhancing the prospects for robust proximal causal inference in the face of hidden confounders and invalid proxies. 
     \item Building on these theoretical results, we illustrate our methods with extensive simulations and an empirical application to the SUPPORT study evaluating the survival causal impact of right-heart catherization.
\end{enumerate}

\section{Notation}
For each individual $i \in\{1, \ldots, n\}$, let $Y_i^{(d)} \in \mathbb{R}$ denote the potential outcome had, possibly contrary to fact, the exposure $D$ been set to $d \in \mathbb{R}$. Suppose that one has observed $n$ i.i.d. realizations of the outcome, treatment, TCP, and OCP.

\vspace{2 mm}

\noindent Let $\mathbf{Y}=\left(Y_1, \ldots, Y_n\right)$ denote the $n$-dimensional vector of observed outcomes, $\mathbf{D}=$ $\left(D_1, \ldots, D_n\right)$ the $n$-dimensional vector of observed treatment variables, 
$\mathbf{Z}$ the $n$ by $p_z$ matrix of candidate TCPs where row $i$ consists of $\mathbf{Z}_{i\cdot}$, and $\mathbf{W}$ denote the $n$ by $p_w$ matrix of OCP variables where row $j$ consists of $\mathbf{W}_{j\cdot}$.

\vspace{2 mm}

\noindent For any vector ${\boldsymbol{\alpha}} \in \mathbb{R}^L$, let $ \alpha_j$ denote the $j$th element of $\boldsymbol{\alpha}$. Let $\|\boldsymbol{\alpha}\|_1,\|\boldsymbol{\alpha}\|_2$, and $\|\boldsymbol{\alpha}\|_{\infty}$ be the usual 1, 2 and $\infty$-norms, respectively. Let $\|\boldsymbol{\alpha}\|_0$ denote the 0-norm, i.e. the number of non-zero elements in $\boldsymbol{\alpha}$. The support of $\boldsymbol{\alpha}$, denoted as $\operatorname{supp}(\boldsymbol{\alpha}) \subseteq\{1, \ldots, L\}$, is defined as the set containing the non-zero elements of the vector $\boldsymbol{\alpha}$, i.e. $j \in \operatorname{supp}(\boldsymbol{\alpha})$ if and only if $\alpha_j \neq 0$. A vector $\boldsymbol{\alpha}$ is called $s$-sparse if it has no more than $s$ non-zero entries. For a vector $\boldsymbol{\alpha} \in \mathbb{R}^L$ and a subset $A \subseteq {1, \ldots, L}$, we use $\boldsymbol{\alpha}_A \in \mathbb{R}^{|A|}$ to denote the subvector of $\boldsymbol{\alpha}$ containing the components indexed by $A$. For an individual index $j \in {1, \ldots, L}$, we write $\alpha_j$ for the $j$-th entry of $\boldsymbol{\alpha}$, and $\boldsymbol{\alpha}_{-j}$ to denote the sub-vector obtained by removing the $j$-th component from $\boldsymbol{\alpha}$.

\vspace{2 mm}

\noindent For any $n$ by $L$ matrix $\mathbf{M} \in \mathbb{R}^{n \times L}$, we denote the $(i, j)$ element of matrix $\mathbf{M}$ as $M_{ij}$, the $i$ th row as $\mathbf{M}_{i .}$, and the $j$ th column as $\mathbf{M}_{\cdot j}$. Let $\mathbf{M}^{\intercal}$ be the transpose of $\mathbf{M}$. Let ${\bf P_{M}}$ be the $n$ by $n$ orthogonal projection matrix onto the column space of $\mathbf{M}$, specifically ${\bf P_{M}}=\mathbf{M}\left(\mathbf{M}^T \mathbf{M}\right)^{-1} \mathbf{M}^{\intercal}$; it is assumed that $\mathbf{M}^{\intercal} \mathbf{M}$ has a proper inverse, unless otherwise noted. Let ${\bf P_{M^{\perp}}}$ be the residual projection matrix, specifically ${\bf P_{M^{\perp}}}=\mathbf{I}-{\bf P_{M}}$ where $\mathbf{I}$ is an $n$ by $n$ identity matrix.
\vspace{2 mm}

\noindent For any sets $A \subseteq\{1, \ldots, L\}$, we denote $A^C$ to be the complement of set $A$. Also, we denote $|A|$ to be the cardinality of set $A$.

\section{Invalid TCP}
\label{sec: invalid z}

To simplify the presentation of the core ideas, we have omitted the measured covariates $\mathbf{X}$ from the theoretical model specification. These covariates may represent additional observed confounders that are not captured by the candidate TCPs $\mathbf{Z}$ or the OCPs $\mathbf{W}$. In practice, we incorporate $\mathbf{X}$ directly alongside the treatment 
$D$ in both identification and estimation. This approach allows for arbitrary associations between $\mathbf{X}$ and the proxies or unmeasured confounder. The real data analysis reflects this adjustment, and our methodology is fully compatible with the inclusion of measured confounders as part of the system.

\vspace{2 mm}
\noindent Consider a point exposure D, outcome Y and an unmeasured confounder U. Suppose, the following simple structural linear model holds :
\begin{equation*}
    \begin{aligned}
        \mathbb{E}(Y \mid D, \mathbf{Z}, U) & = \beta D + \boldsymbol{\alpha}^{\intercal}\mathbf{Z} + \beta_u U, \\
        \mathbb{E}(W \mid D, \mathbf{Z}, U) & = \eta_u U,
    \end{aligned}
\end{equation*}
where $\mathbf{Z} = (Z_{i1}, \ldots, Z_{ip_z})^{\intercal}$ is the vector of candidate TCPs, some of which may be invalid in the sense that they may impact $Y$ (but not $W$). With $\eta_u \neq 0$, $W$ serves as a valid OCP because it is related to $D$ and $\mathbf{Z}$ 
solely through $U$, and not through any direct associations. It is straightforward to show that
\begin{equation}
    \mathbb{E}(Y \mid D,Z) = \beta D + \boldsymbol{\alpha}^{\intercal}\mathbf{Z} + \gamma \mathbb{E}(W \mid D, Z),
    \label{eq: fmodel}
\end{equation}
where $\gamma = \frac{\beta_{u}}{\eta_{u}}$. 
If we have a random sample $\{Y_i, D_i, \mathbf{Z}_{i\cdot}, \mathbf{W}_i\}_{i=1}^{n}$ then \eqref{eq: fmodel} can be equivalently written as 
\begin{equation}
    Y_i = D_i \beta + \mathbf{Z}_{i\cdot}^{\intercal}\boldsymbol{\alpha} + W_i\gamma + \epsilon_i; \quad 1\leq i \leq n.
    \label{eq: model}
\end{equation}
where $\E(\epsilon \mid D, \mathbf{Z}) = 0$. Further assume that $\E(\epsilon^2 \mid D, \mathbf{Z}) = \sigma_{\epsilon}^2$. Based on this observed model in \eqref{eq: model}, the parameter $\alpha$ represents the direct effect of the TCPs 
on the outcome. If there is no direct effect of the $j$-th TCP then $\alpha_j = 0$. 
Hence the value of $\boldsymbol{\alpha}$ directly encodes which TCP is valid and which is invalid. In the following we formalize the definition of a valid proxy.

\vspace{2 mm}
\noindent {\bf Definition 1:} Suppose we have model \eqref{eq: model} with $p$ treatment inducing proxies. We say proxy $j \in \{1,2,\ldots,p\}$ is valid if $\alpha_j = 0$ and invalid if $\alpha_j \neq 0$.



\subsection{Oracle Estimator for the Causal Effect}
\label{sec: oracle Z}

Let $A := \{j : \alpha_j \neq 0\}$ represent the set of invalid TCPs and $s_z:= |A|$ represent the number of invalid TCPs. The oracle P2SLS estimator is obtained when $\mathbf{Z}_A$—the set of invalid proxies—is known and correctly treated as measured potential confounders rather than as proxies, since their direct causal effect on the outcome $Y$ violates the exclusion restriction required for valid TCPs. The oracle estimator $\widehat{\beta}_{or}$ is then computed in two steps:
\begin{enumerate}
    \item[] {\bf Step 1.} Regress $\mathbf{W}$ on $\mathbf{M} := (\mathbf{Z}, \mathbf{D})$. $\mathbf{\widehat{W}} = P_M \mathbf{W}$.
    \item[] {\bf Step 2.} Regress $\mathbf{Y}$ on $\mathbf{D}, \mathbf{Z}_{A}, \widehat{\mathbf{W}}$. The parameters are estimated via least squares of the model $Y_i = D_i\beta + \mathbf{Z}_{iA}\boldsymbol{\alpha}_{A} + \widehat{W}_i\gamma + \xi_i$ where $\xi_i$ is defined implicitly.
\end{enumerate}
Therefore,
$$
\widehat{\beta}_{or} = \frac{\mathbf{D}^{\intercal}\mathbf{P_{\widehat{N}_A^{\perp}}}\mathbf{Y}}{\mathbf{D}^{\intercal}\mathbf{P_{\widehat{N}_A^{\perp}}}\mathbf{D}}
\label{eq: oracle Z}
$$
where $\mathbf{N}_A := (\mathbf{Z}_A \quad \mathbf{W})$ and $\widehat{\mathbf{N}}_A := \mathbf{P_M} \mathbf{N}_A$ is the projection of the invalid TCPs and $W$ onto the columnspace of $\mathbf{M}$. Under standard assumptions it can be proved that
\begin{equation}
    \sqrt{n}(\widehat{\beta}_{or} - \beta) \to_d N(0, \sigma^2_{or})
\end{equation}
where $\sigma^2_{or}$ is given in Appendix \ref{app:oracle_variance}. 

\subsection{Identifiability with Some Invalid Proxies}
We are now ready to consider the more challenging setting in which the set of invalid proxies $\mathbf{Z}_A$ is not known a priori. In this vein, recall $\mathbf{M} := (\mathbf{Z}, \mathbf{D})$. We begin by making the following assumptions:
\begin{enumerate}
    \item \label{ass a}
    $\mathbb{E}(\mathbf{M}^{\intercal}\mathbf{M})$ is full rank. 
    \item \label{ass b} 
    For $\delta^* := \mathbb{E}(\mathbf{M}^{\intercal}\mathbf{M})^{-1}\E(\mathbf{M}^{\intercal}\mathbf{W}), \delta^*_j \neq 0$ for $1 \leq j \leq (p_z+1)$. 
    \item  \label{ass c}
    The number of invalid TCPs, $s_z$, must be less than some number $I$, i.e., $s_z < I$, without knowing which TCPs are invalid or knowing the exact number of invalid TCPs.
\end{enumerate}
Now the model in \eqref{eq: model} implies the moment condition
\begin{equation}
    \mathbb{E}\left(\mathbf{M}^{\intercal}(\mathbf{Y} - \mathbf{D}\beta - \mathbf{Z}\boldsymbol{\alpha} - \mathbf{W}\gamma)\right) = 0
    \label{eq: moment}
\end{equation}
Under assumption \ref{ass a}, \eqref{eq: moment} simplifies to
\begin{equation}
    \begin{aligned}
        \mathbf{\Gamma}^*_{-(p_z+1)} & = \boldsymbol{\alpha} + \boldsymbol{\delta}^{*}_{-(p_z+1)}\gamma \\
        \Gamma^*_{p_z+1} & = \beta + \delta^{*}_{p_z+1}\gamma
    \end{aligned}
    \label{eq: simple moment}
\end{equation}
where $\mathbf{\Gamma}^* := \mathbb{E}(\mathbf{M}^{\intercal}\mathbf{M})^{-1}\mathbb{E}(\mathbf{M}^{\intercal}\mathbf{Y})$ and $\boldsymbol{\delta}^{*} := \mathbb{E}(\mathbf{M}^{\intercal}\mathbf{M})^{-1}\mathbb{E}(\mathbf{M}^{\intercal}\mathbf{W})$ are identifiable. If we can identify $\boldsymbol{\alpha}$ and $\gamma$ from the first set of equations $\mathbf{\Gamma}^*_{-(p_z+1)} = \boldsymbol{\alpha} + \boldsymbol{\delta}^{*}_{-(p_z+1)}\gamma$, then $\beta$ can be uniquely identified from the last equation as $\Gamma^*_{p_z+1} - \delta_{p_z+1}^{*}\gamma$. The following theorem provides the necessary and sufficient conditions for identifying $\alpha$ and $\gamma$, and consequently $\beta$. This result follows directly from Theorem 1 in \cite{kang2016}. 

\begin{Theorem}
Consider the model \eqref{eq: model} under assumptions \ref{ass a}, \ref{ass b} and \ref{ass c}. 
For an upper bound $I$ for $s_z$ with $I, s_z \in \{1,2,\ldots,p_z\}$ and $s_z < I$, 
consider all sets $C_m \subset \{1,\ldots,p_z\}$, $m = 1,\ldots, M_1$ of size $|C_m| =p_z- I + 1$ such that $\widetilde{\boldsymbol{\delta}}_{j\cdot}q_m = \widetilde{\boldsymbol{\Gamma}}_j, j \in C_m$ where $q_m$ is a constant, $\widetilde{\boldsymbol{\delta}} := \boldsymbol{\delta}^*_{-(p_z+1)}, \widetilde{\boldsymbol{\Gamma}} := \boldsymbol{\Gamma}^*_{-(p_z+1)}$. There is a unique solution $\boldsymbol{\alpha}$ and $\gamma$ to \eqref{eq: simple moment} if and only if $q_1=q_2=\dots=q_{M_1}$.
\label{thm: identification}
\end{Theorem}

\vspace{2mm}
\noindent
We now illustrate the conditions of Theorem~\ref{thm: identification} through two concrete examples both with $p_z=3$.
\begin{itemize}
\item[] \textbf{Example 1.} Let $\widetilde{\boldsymbol{\delta}} = (1,2,3,4)$, $\widetilde{\boldsymbol{\Gamma}} = (1,2,3,8)$, and $I = 3$. Then, there are $M_1=3$ possible subsets of size $p_z-I+1=2$, which are $C_1 = \{1,2\}$, $C_2 = \{1,3\}$, and $C_3 = \{2,3\}$. The three subsets each satisfy $\widetilde{\boldsymbol{\delta}}_{j\cdot} q_m = \widetilde{\boldsymbol{\Gamma}}_j$ with $q_1 = q_2 = q_3 = 1$. Since all $q_m$ are equal, the consistency condition is satisfied, and a unique solution for $(\boldsymbol{\alpha}, \gamma)$ exists.


    \vspace{1mm}
    \item[] \textbf{Example 2.} Let $\widetilde{\boldsymbol{\delta}} = (1,2,3,4)$, $\widetilde{\boldsymbol{\Gamma}} = (1,2,6,8)$. Here, the sets $C_1 = \{1,2\}$ with $q_1 = 1$ and $C_2 = \{3,4\}$ with $q_2 = 2$ satisfy $\widetilde{\boldsymbol{\delta}}_{j\cdot} q_m = \widetilde{\boldsymbol{\Gamma}}_j$ but $q_1 \ne q_2$. Hence, the condition of the theorem fails and a unique solution does not exist.
\end{itemize}
Theorem \ref{thm: identification} states that $ \gamma $ is identifiable if there are no two subsets of TCPs of size $p_z- I + 1 $ that give consistent estimates of $ \gamma $ within each subset, but produce different estimates when compared across subsets. 
However, verifying this condition can be computationally intensive, especially when $ I $ is large. This is because checking all possible subsets of size $p_z- I + 1 $ from the set of $p_z$ proxies requires evaluating a large number of combinations, specifically $ \binom{p_z}{p_z - I + 1} $, and considering the associated constants $ q_m $ from $\widetilde{\boldsymbol{\Gamma}} $ and $ \widetilde{\boldsymbol{\delta}} $. Corollary \ref{corr: majority} below provides a more straightforward sufficient identification condition which ensures that the consistency condition automatically holds when $ I \leq p_z/2 $,  and implies that at most half of the proxies are invalid. This eliminates the need for an exhaustive subset check, making the identification process more practically manageable.
\begin{Corollary}
    If $I \leq \frac{p_z}{2}$, there is always a unique solution to \eqref{eq: simple moment}.
    \label{corr: majority}
\end{Corollary}
In addition to its computational advantages, Corollary 1 offers a simpler interpretation compared to Theorem 1. For instance, consider a situation where an analyst is working with a set of TCPs but does not have complete knowledge about which proxies are valid or invalid. As long as fewer than $50\%$ of the total proxies are invalid, the analyst can be confident that the parameters will always be identifiable, without having to identify each proxy's validity individually. This simplifies the analysis, particularly in cases where the validity of all proxies is particularly challenging to determine, however, the analyst is quite confident that the majority of proxies are valid. In practice, we typically rely on this majority rule for simplicity, and we adopt this assumption in the rest of the paper to streamline the discussion, even though Theorem \ref{thm: identification} allows for a weaker condition.
\subsection{Estimation of the Causal Effect}

Motivated by the moment condition \eqref{eq: moment}, we estimate the parameters $(\boldsymbol{\alpha}^{\intercal},\beta,\gamma)^{\intercal}$ by the following optimization
\begin{equation}
    \left(\widehat{\beta},\widehat{\boldsymbol{\alpha}},\widehat{\gamma}\right)  = \arg\min_{\beta,\boldsymbol{\alpha},\gamma}\left\|\mathbf{P_{M}}(\mathbf{Y} - \mathbf{D}\beta - \mathbf{Z}\boldsymbol{\alpha} - \mathbf{W}\gamma)\right\|_2^2 + \lambda\|\boldsymbol{\alpha}\|_1
    \label{eq: LASSO}
\end{equation}
where $\lambda > 0$ is some tuning parameter. 
While the objective in \eqref{eq: LASSO} resembles that of the traditional LASSO 
our use of the $\ell_1$-penalty serves a different purpose. Unlike standard LASSO, which typically penalizes all regression coefficients and is aimed at improving predictive accuracy, we penalize only the coefficients $\boldsymbol{\alpha}$ associated with the TCPs $\mathbf{Z}$. Our goal is not prediction but rather bias reduction of the unpenalized treatment effect $\beta$, by encouraging sparsity in $\boldsymbol{\alpha}$ and thereby selecting valid proxies whose coefficients are zero. The following theorem provides an alternative representation of the estimator in \eqref{eq: LASSO} that allows for a convenient two-step implementation and yields the same solution. 

\begin{Theorem}
Define $\widehat{\mathbf{W}} = \mathbf{P_{M}} \mathbf{W}$. With $\widetilde{\mathbf{D}} := \mathbf{P_{\widehat{W}^{\perp}}D}$ and $\widetilde{\mathbf{Z}} := \mathbf{P_{\widehat{W}^{\perp}}Z}$, we propose the following two-step algorithm
\begin{equation}
    \begin{aligned}
        \widehat{\boldsymbol{\alpha}} & = 
        \arg\min_{\alpha}\frac{1}{2}\left\|\mathbf{Y} - \mathbf{P_{\widetilde{D}^{\perp}}\widetilde{\mathbf{Z}}}\boldsymbol{\alpha}\right\|_2^2 + \lambda\|\boldsymbol{\alpha}\|_1 \\
        \widehat{\beta} & = \frac{\widetilde{\mathbf{D}}^{\intercal}(\mathbf{Y} - \widetilde{\mathbf{Z}}\widehat{\boldsymbol{\alpha}})}{\|\widetilde{\mathbf{D}}\|_2^2}.
    \end{aligned}
\label{eq: LASSO 2 step}
\end{equation}
\label{thm: LASSO 2 step}
\end{Theorem}
This formulation offers a convenient two-step implementation: we first estimate the nuisance component $\boldsymbol{\alpha}$ using a Lasso regression of $\mathbf{Y}$ on the orthogonalized proxies $\mathbf{P_{\widetilde{D}^{\perp}}}\widetilde{\mathbf{Z}}$, and then estimate $\beta$ by regressing the residuals on the orthogonalized treatment $\widetilde{\mathbf{D}}$. Notably, if we replace $\widehat{\boldsymbol{\alpha}}$ with the true $\boldsymbol{\alpha}$ in the expression for $\widehat{\beta}$, we recover the oracle estimator $\widehat{\beta}_{\mathrm{or}}$, since
$$
\frac{\widetilde{\mathbf{D}}^{\intercal}(\mathbf{Y} - \widetilde{\mathbf{Z}}\boldsymbol{\alpha})}{\|\widetilde{\mathbf{D}}\|_2^2} = \frac{\widetilde{\mathbf{D}}^{\intercal}(\mathbf{Y} - \widetilde{\mathbf{Z}}_A\boldsymbol{\alpha}_A)}{\|\widetilde{\mathbf{D}}\|_2^2},
$$
and by the normal equations, this expression corresponds to the OLS coefficient on $\mathbf{D}$ from regressing $\mathbf{Y}$ on $(\mathbf{D}, \mathbf{Z}_A, \widehat{\mathbf{W}})$, which matches the oracle causal effect estimator in \eqref{eq: oracle Z}.
\subsection{Theoretical Guarantee}
In this Section, we provide a theoretical analysis of the LASSO-based proximal estimator of the previous Section. First, we require some key definitions.

\vspace{2 mm}
\noindent{\bf Definition 2:} For any matrix $M$, the upper and lower restricted isometry property (RIP) constants of order $k$, denoted as $\delta_k^{+}(\mathbf{M})$ and $\delta_k^{-}(\mathbf{M})$ respectively, are the smallest $\delta_k^{+}(\mathbf{M})$ and largest $\delta_k^{-}(\mathbf{M})$ such that
$$
\delta_k^{-}(\mathbf{M})\|\boldsymbol{\alpha}\|_2^2 \leq\|\mathbf{M} \boldsymbol{\alpha}\|_2^2 \leq \delta_k^{+}(\mathbf{M})\|\boldsymbol{\alpha}\|_2^2
$$
holds for all $k$-sparse vectors $\boldsymbol{\alpha}$.

\vspace{2 mm}
\noindent The RIP has been applied in high-dimensional linear regression to establish the consistency and performance of various estimators, including the LASSO and Dantzig selector (Candes and Tao, 2007; Bickel et al., 2009; Wainwright, 2009), and plays a key role in establishing sharp recovery thresholds and theoretical guarantees for sparse regression methods (Meinshausen and Bühlmann, 2010; van de Geer et al., 2014).

\vspace{2 mm}
\begin{Theorem}
Let the restricted isometry property constants satisfy the following
$$
2\delta^{-}_{2s_z}(\mathbf{Z}) > \delta^{+}_{2s_z}(\mathbf{Z}) + 2\delta^{+}_{2s_z}(\mathbf{P_{\widehat{W}}Z}) + 2\delta^{+}_{2s_z}(\mathbf{P_{\widetilde{D}}Z})
$$
where $s_z := |A|$ is the number of invalid TCPs. Then if $\lambda \geq 3 \left\|\mathbf{Z}^{\intercal}\mathbf{P_{\widehat{W}^{\perp}}P_{\widetilde{D}^{\perp}}P_M\epsilon}\right\|_{\infty}$,
\begin{equation}
\begin{aligned}
    \|\mathbf{h}_{A^c}\|_1 & \leq 2 \|\mathbf{h}_A\|_1 \\
    \|\mathbf{h}_A\|_2 & \leq \frac{4/3\lambda\sqrt{s_z}}{2\delta^{-}_{2s_z}(\mathbf{Z}) - \delta^{+}_{2s_z}(\mathbf{Z}) - 2\delta^{+}_{2s_z}(\mathbf{P_{\widehat{W}}Z}) - 2\delta^{+}_{2s_z}(\mathbf{P_{\widetilde{D}}Z})} 
\end{aligned} 
\label{eq: alpha guarantee}
\end{equation}
where $h := \widehat{\boldsymbol{\alpha}} - \boldsymbol{\alpha}$ and $A := \{j: \alpha_j \neq 0\}$ is the set of invalid TCPs.
\label{thm: alpha guarantee}    
\end{Theorem}
The proof of Theorem 3 follows similarly to the proof of Theorem 2 in \cite{kang2016} and is therefore omitted here to avoid redundancy. Theorem \ref{thm: alpha guarantee} provides conditions for the reliable estimation of the parameter $\boldsymbol{\alpha} $ when using regularization. It shows that if the restricted isometry constants satisfy certain inequalities and the regularization parameter $ \lambda $ is appropriately chosen, the estimator $ \widehat{\boldsymbol\alpha} $ will provide accurate estimates. Specifically, the error $ \boldsymbol h = \widehat{\boldsymbol\alpha} - \boldsymbol\alpha $ is bounded in both $ \ell_1 $ and $ \ell_2 $ norms, with the bounds depending on the sparsity level $s_z$ of the true parameter vector $\boldsymbol \alpha$ — that is, the number of invalid TCPs—and the restricted isometry properties of the TCP matrix $Z$ and its projections. This result is crucial for ensuring the robustness of the estimator in the presence of invalid TCPs.

\vspace{3 mm}
\noindent To better understand the implications of Theorem \ref{thm: alpha guarantee}, consider a regime where the number of candidate proxies $p_z$ is moderate and the number of invalid TCPs $s_z$ remains relatively small. Suppose the TCP matrix $\mathbf{Z}$ satisfies standard regularity conditions — specifically, that the restricted isometry constants are bounded as $\delta^{-}_{2s_z}(\mathbf{Z}) \geq c_1$ and $\delta^{+}_{2s_z}(\mathbf{Z}) \leq c_2$, with constants $c_1, c_2 > 0$. Further, the projection terms $\mathbf{P_{\widehat{W}}Z}$ and $\mathbf{P_{\widetilde{D}}Z}$ have restricted isometry constants bounded above by a small constant $c_3 > 0$. This happens, for example, when:
\begin{itemize}
    \item 

    The invalid TCPs (nonzero entries of  $\boldsymbol{\alpha}^*$ ) are only weakly associated with the variation in  $\widehat{W} = \mathbf{P}_M W$. While  $W$  is a valid OCP linked to the hidden confounder  $U$ ,  $\widehat{W}$  is the part of  $W$  explained by observed data—namely treatment  $D$  and candidate TCPs $\mathbf{Z}$ . For $\delta^{+}_{2s_z}(\mathbf{P}_{\widehat{W}} \mathbf{Z})$ to be small, invalid TCPs must not closely mimic  $\widehat{W}$  or carry the same confounding signal after accounting for  $D$  and  $\mathbf{Z}$ . If they did, they would align strongly with  $\widehat{W}$ , increasing the RIP constant and violating the condition. Thus, invalid TCPs should have weaker or different links to  $U$  than the part of  $W$  captured by $\widehat{W}$.
    
    
    \item The invalid TCPs (nonzero entries of $\boldsymbol{\alpha}^*$) are weakly associated with the residualized treatment $\widetilde{D} = (\mathbf{I} - \mathbf{P_{\widehat{W}}}) D$, which represents the part of $D$ unexplained by $\widehat{W} = \mathbf{P_M} W$. In the DAG, $Z_2$ affects $D$ both directly and via $U$, making it potentially predictive of $\widetilde{D}$. To ensure $\delta^{+}_{2s_z}(\mathbf{P_{\widetilde{D}} Z})$ remains small, these pathways must be weak so that invalid TCPs do not explain the variation in $D$ left out by $\widehat{W}$. 
\end{itemize}
Under these conditions, the denominator in the $\ell_2$-bound simplifies to a positive constant $c > 0$, and we obtain
$$
\|\mathbf{h}_A\|_2 \lesssim \lambda \sqrt{s_z}.
$$
Now suppose the error term $\epsilon$ has bounded sub-Gaussian tails (e.g., mean zero with variance proxy $\sigma^2$). Then with high probability, the quantity $\|\mathbf{Z}^\intercal \mathbf{P}_{\widehat{W}^\perp} \mathbf{P}_{\widetilde{D}^\perp} \mathbf{P}_M \boldsymbol\epsilon\|_\infty$ will scale as
$$
O_{\mathbb{P}}\left( \sigma \sqrt{\frac{\log p_z}{n}} \right).
$$
Thus, setting the regularization parameter $\lambda \asymp \sigma \sqrt{\frac{\log p_z}{n}}$ ensures that the condition in Theorem \ref{thm: alpha guarantee} holds. Substituting this into the bound yields the overall error rate
$$
\|\widehat{\boldsymbol{\alpha}} - \boldsymbol{\alpha}\|_2 \lesssim \sigma \sqrt{\frac{s_z \log p_z}{n}}.
$$
This rate guarantees consistency of the estimator as long as $s_z \log p_z = o(n)$, a mild sparsity condition often met in applications where most proxies are valid.

\subsection{Using Adaptive LASSO for Estimation of the Causal Effect}
\label{sec: adaptive LASSO}

The results in \cite{windmeijer2019} suggest that the LASSO path may fail to select the correct model, leading to an inconsistent estimator of $\beta$, a challenge we expect to encounter in our proximal setup as well. This issue persists even under the majority rule, where fewer than $50\%$ of the proxies are invalid, due to the correlation patterns among them. Specifically, the irrepresentable condition, as defined by  \cite{yuzhao, zhou2006}, 
\begin{equation}
\left\|\mathbf{C}_{A^c A} \mathbf{C}_{A A}^{-1} \mathbf{s}\left(\boldsymbol{\alpha}_A\right)\right\|_{\infty} < 1,
\end{equation}
is a necessary condition for LASSO variable selection consistency. Here, $\mathbf{C} = \text{plim} \frac{1}{n} (\mathbf{P_{\widetilde{D}^{\perp}}} \widetilde{\mathbf{Z}})^\intercal (\mathbf{P_{\widetilde{D}^{\perp}}} \widetilde{\mathbf{Z}})$ and $\mathbf{C}$ is partitioned as 
$$
\begin{pmatrix}
    \mathbf{C}_{AA} & \mathbf{C}_{AA^c} \\
    \mathbf{C}_{AA^c}^\intercal & \mathbf{C}_{A^cA^c}
\end{pmatrix}
$$
where $\mathbf{C}_{AA}$ is an $s_z \times s_z$ matrix. 

\vspace{2 mm}
\noindent Mathematically, the irrepresentable condition ensures that the correlation between the covariates in the active set $A$ (those with non-zero coefficients) and the covariates outside this set $A^c$(those whose coefficients are zero) is sufficiently low. This allows LASSO to correctly distinguish between relevant and irrelevant variables. The expression $\left\|\mathbf{C}_{A^c A} \mathbf{C}_{A A}^{-1} \mathbf{s}(\boldsymbol{\alpha}_A)\right\|_{\infty} < 1$ checks that the correlation between the irrelevant variables and the active variables is small enough for LASSO to select only the active ones correctly. In our context, the design matrix is $\mathbf{P_{\widetilde{D}^{\perp}} \mathbf{P_{\widehat{W}^{\perp}}} Z} = \mathbf{P_{\widetilde{D}^{\perp}}} Z$, effectively removing the indirect effects of $W$ and $D$ on $Z$. 

\begin{figure}[H]
        \centering
        \scalebox{0.95}{ 
        \begin{tikzpicture}[node distance={20mm}, main/.style = {draw, circle}, rednode/.style = {draw=red, circle, thick}] 
            \node[main] (1) {$U$};
            \node[main] (2) [below left of = 1] {$Z_1$};
            \node[main] (3) [below right of = 1] {$W$};
            \node[main] (4) [below of = 2] {$D$};
            \node[main] (5) [below of = 3] {$Y$};
            \node[rednode] (6) [below of = 4] {$Z_2$};
            \draw[->, >=latex, line width=0.5mm] (1) -- (2);
            \draw[->, >=latex, line width=0.5mm] (1) -- (3);
            \draw[->, >=latex, line width=0.5mm] (2) -- (4);
            \draw[->, >=latex, line width=0.5mm] (3) -- (5);
            \draw[->, >=latex, line width=0.5mm] (4) -- (5);
            \draw[->, >=latex, line width=0.5mm] (1) -- (4);
            \draw[->, >=latex, line width=0.5mm] (1) -- (5);
            \draw[->, >=latex, line width=0.5mm] (6) -- (4);
            \draw[->, >=latex, line width=0.5mm, red] (6) -- (5);
            \draw[->, >=latex, line width=0.5mm, bend right=65] (1) to (6);
            \draw[->, >=latex, line width=0.5mm, blue, bend right=40] (2) to (6);
        \end{tikzpicture}
        }
        \caption{\small $Z_2$ is an invalid TCP, $Z_1$ affects $Z_2$.}
        \label{fig: violation irr}
    \end{figure}
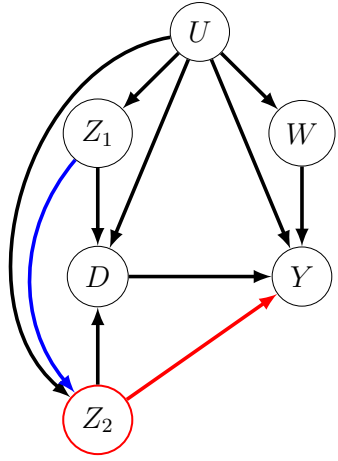

\vspace{2 mm}
\noindent Thus, the irrepresentable condition requires that the projected covariates—specifically, $\mathbf{P_{\widetilde{D}^{\perp}} Z_{\cdot A}}$ and $\mathbf{P_{\widetilde{D}^{\perp}} Z_{\cdot A^c}}$ —are not highly correlated. When this condition fails, LASSO struggles to distinguish between relevant and irrelevant variables, leading to incorrect variable selection. Figure \ref{fig: violation irr} illustrates such a violation. Here, $Z_1$ is a valid TCP (with no direct effect on $Y$), while $Z_2$ is an invalid TCP, affecting $Y$ directly (red arrow). Crucially, the blue arrow indicates a direct causal path from $Z_1$ to $Z_2$, which induces strong correlation between them—even after adjusting for the effects of $D$ and $W$ through projection. As a result, LASSO may either assign non-zero coefficients to both $Z_1$ and $Z_2$, or shrink both towards zero, making it impossible to reliably identify which TCP is truly valid.

\noindent To address these challenges arising from violations of the irrepresentable condition, the adaptive LASSO offers a more reliable approach with better variable selection properties. As described below, we employ a $\sqrt{n}$-consistent initial estimator for $\alpha$ in \eqref{eq: median alpha}, followed by the adaptive LASSO algorithm to estimate the parameters:
\begin{equation}
    \begin{aligned}
        \widehat{\boldsymbol\alpha}_{ad} & = \arg\min_{\boldsymbol\alpha,\gamma} \left\|\bf Y - P_{\widetilde{D}^{\perp}}\widetilde{\mathbf{Z}}\boldsymbol\alpha \right\|_2^2 +  \lambda_n\sum_{i=1}^{p_z}\frac{|\alpha_{i}|}{|\widehat{\alpha}^m_{i}|}, \\
        \widehat{\beta}_{a d} & = \frac{\widetilde{\mathbf{D}}^{\intercal}(\bf Y - \widetilde{\mathbf{Z}}\widehat{\boldsymbol\alpha}_{ad})}{\|\widetilde{\mathbf{D}}\|_2^2} \\
    \end{aligned}
    \label{eq: adaptive LASSO}
\end{equation}
This approach improves variable selection by assigning smaller penalties to variables with strong initial estimates and larger penalties to those with weaker signals, helping to more accurately distinguish relevant variables from irrelevant ones.

\vspace{2 mm}

\noindent\underline{{\bf Consistent Estimator for $\boldsymbol \alpha$}}

\vspace{3 mm}

\noindent In the standard linear model, the OLS estimator remains consistent when all explanatory variables are included. However, $\bf P_{\widetilde{D}^{\perp}}\widetilde{\mathbf{Z}}$ is not full rank, meaning that we cannot set $\lambda_n = 0$ in \eqref{eq: adaptive LASSO}. Consequently, we follow the methodology from \cite{windmeijer2019} to construct a consistent estimator for $\boldsymbol \alpha$, assuming the majority rule holds.

\vspace{2 mm}

\noindent Define $\widehat{\mathbf{\Gamma}} := (\mathbf{M}^{\intercal}\mathbf{M})^{-1}\mathbf{M}^{\intercal}\mathbf{Y}$ and $\widehat{\boldsymbol \delta} := (\mathbf{M}^{\intercal}\mathbf{M})^{-1}\mathbf{M}^{\intercal}\mathbf{W}$. Estimate $\gamma$ by $\widehat{\gamma}^m := \operatorname{median}(\widehat{\boldsymbol{\pi}})$ where for $1 \leq j \leq p_z$,
$$
\widehat{\pi}_j = \frac{\widehat{\Gamma}_j}{\widehat{\delta}_j}
$$

\begin{lemma}
Under model specification \eqref{eq: model} along with assumptions \ref{ass a} and \ref{ass b}, if $s_z <p_z/ 2$, then the estimator $\widehat{\gamma}^m$ defined as
$$
\widehat{\gamma}^m=\operatorname{median}(\widehat{\boldsymbol{\pi}})
$$
is a consistent estimator for $\gamma$,
$$
\operatorname{plim}\left(\widehat{\gamma}^m\right)=\gamma
$$
The limiting distribution of $\widehat{\gamma}^m$ is given by
$$
\sqrt{n}\left(\widehat{\gamma}^m-\gamma\right) \xrightarrow{d} q_{[l], p_z-s_z},
$$
where for $p_z$ odd, $q_{[l], p_z-s_z}$ is the $l$ th-order statistic of the limiting normal distribution of $\sqrt{n}\left(\widehat{\boldsymbol \pi}_{A^c}-\gamma \mathbf{1}_{p_z-s_z}\right)$, where $l$ is determined by $p_z, s_z$, and the signs of $\frac{\alpha_j}{\delta_j}, j=1, \ldots, s_z$. For $p_z$ even, $q_{[l], p_z-s_z}$ is defined as the average of either the $l$ and $[l-1]$-order statistics, or the $[l]$ and $[l+1]$-order statistics.
\label{lem: median est}
\end{lemma}

\noindent Finally estimate $\boldsymbol \alpha$ by 
\begin{equation}
    \widehat{\boldsymbol \alpha}^m := \widehat{\mathbf{\Gamma}}_{-(p_z+1)} - \widehat{\gamma}_m\widehat{\boldsymbol\delta}_{-(p_z+1)}.
    \label{eq: median alpha}
\end{equation}
\noindent The following proposition summarizes the properties of the estimator in \eqref{eq: adaptive LASSO} and and directly follows from the theoretical results of the adaptive LASSO estimator established in \cite{zhou2006}.
\begin{Proposition}
    Suppose $\lambda_n = o(\sqrt{n})$ and $\lambda_n \rightarrow \infty$, then the adaptive LASSO estimator $\widehat{\alpha}_{a d}$ satisfies
    \begin{enumerate}
        \item Consistency in variable selection : $\lim_{n \rightarrow \infty}P(\widehat{A}_{a d} = A) = 1$ where $\widehat{A}_{a d} = \{j : \widehat{\alpha}_{a d,j} \neq 0\}$ and $A = \{j : \alpha_{j} \neq 0\}$.

        \item Asymptotic normality : $\sqrt{n}(\widehat{\boldsymbol\alpha}_{ad,A}-\boldsymbol\alpha_{A}) \rightarrow^{d} N(0,\sigma^2\mathbf{C}_{AA}^{-1}) $.
    \end{enumerate}
    \label{prop: adaptive LASSO alpha g}
\end{Proposition}
Finally, regress $\mathbf{Y}$ on $\mathbf{D}$, the invalid TCPs selected by the adaptive LASSO $\mathbf{Z}_{\widehat{A}_{ad}}$, and $\widehat{\mathbf{W}}$. The post-adaptive LASSO parameter estimates $\widehat{\beta}_{post}$, $\widehat{\boldsymbol{\alpha}}_{ad}$, and $\widehat{\gamma}_{ad}$ are obtained as the ordinary least squares (OLS) estimates from the model

$$
Y_i = D_i \beta + \mathbf{Z}_{i \widehat{A}_{ad}} \boldsymbol{\alpha}_{ad} + \widehat{W}_i \gamma + \xi_i,
$$
where $\xi_i$ is the error term. Here the corresponding parameter estimates $\widehat{\boldsymbol{\alpha}}_{ad}$, $\widehat{\gamma}_{ad}$, and $\widehat{\beta}_{post}$ are obtained via OLS in this regression. Consequently, the post-adaptive LASSO two-stage least squares (2SLS) estimator for the causal effect $\beta$ is given by

\begin{equation}
\widehat{\beta}_{post} := \frac{\mathbf{D}^\intercal \mathbf{P_{\widehat{N}_{ad}^\perp}} \mathbf{Y}}{\mathbf{D}^\intercal \mathbf{P_{\widehat{N}_{ad}^\perp}} \mathbf{D}},
\label{eq: post-ad LASSO}
\end{equation}
where $\widehat{\mathbf{N}}_{ad} := \left(\mathbf{Z}_{\widehat{A}_{ad}} \quad \widehat{\mathbf{W}}\right)$. The variance of $\widehat{\beta}_{post}$ is estimated by $\widehat{\sigma}^2$, which is constructed by replacing each component of the oracle variance $\sigma^2_{or}$ (detailed in the Appendix) with its sample analog. In particular, the error variance $\sigma_{\epsilon}^2$ is approximated by
$\widehat{\sigma}_{\epsilon}^2 = \frac{1}{n} \widehat{\boldsymbol{\epsilon}}^\intercal \widehat{\boldsymbol{\epsilon}},
$ where the residual vector $\widehat{\boldsymbol{\epsilon}}$ is $\widehat{\boldsymbol{\epsilon}} = \mathbf{Y} - \mathbf{D} \widehat{\beta}_{post} - \mathbf{Z}_{\widehat{A}_{ad}} \widehat{\boldsymbol{\alpha}}_{ad} - \widehat{\mathbf{W}} \widehat{\gamma}_{ad}$. Based on this variance estimate, a $100(1-\alpha)\%$ confidence interval for $\beta$ can be constructed as

\begin{equation}
\text{CI}_{\alpha}(\beta) = \left[\widehat{\beta}_{post} - z_{\alpha/2} \widehat{\sigma}, \quad \widehat{\beta}_{post} + z_{\alpha/2} \widehat{\sigma}\right],
\label{eq: CI Z}
\end{equation}
where $z_{\alpha/2}$ is the upper $\alpha/2$-quantile of the standard normal distribution. Table \ref{tab: algo inv Z} outlines the estimation procedure when invalid candidate TCPs are present alongside one valid OCP.

    \begin{table}[H]
    \begin{tabular}{|l|}
    \hline
    {\bf Step 1.} $\widehat{\mathbf{W}} = \mathbf{P_M W}$ \\
    \\
    {\bf Step 2.} Construct the initial estimator $\widehat{\boldsymbol \alpha}^m$ as defined in \eqref{eq: median alpha}, and use it to derive the adaptive LASSO \\
    \hspace{35 pt} estimator $\widehat{\boldsymbol \alpha}_{ad}$, as given in \eqref{eq: adaptive LASSO}.\\
    \\
    {\bf Step 3.} Estimate the set of invalid proxies as $\widehat{A}_{ad} := \{j : \widehat{\alpha}_{ad, j} \neq 0\}$.\\
    \\
    {\bf Step 4.} Finally, regress $\mathbf{Y}$ on $\mathbf{D}, \mathbf{Z}_{\widehat{A}_{ad}}$, and $\widehat{\mathbf{W}}$. Specifically, estimate $\beta$ using the post-adaptive LASSO \\
    \hspace{35 pt} estimator $\widehat{\beta}_{post}$, as defined in \eqref{eq: post-ad LASSO}.\\
    \\
    \hline
    \end{tabular}
    \caption{Estimation Algorithm in Presence of Invalid TCPs}
    \label{tab: algo inv Z}
    \end{table}

\vspace{2 mm}
\noindent Proposition \ref{prop: adaptive LASSO alpha g} leads us to the following proposition.
\begin{Proposition}
Under assumptions \ref{ass a} and \ref{ass b}, the majority validity condition for the TCPs and the conditions of Proposition \ref{prop: adaptive LASSO alpha g}, the limiting distribution of the post-adaptive LASSO estimator $\widehat{\beta}_{post}$ is same as that of the oracle estimator $\widehat{\beta}_{or}$. Specifically,
\begin{equation*}
    \sqrt{n}\left(\widehat{\beta}_{post}-\beta\right) \stackrel{d}{\longrightarrow} N\left(0, \sigma_{or}^2\right),
\end{equation*}
where $\sigma_{or}^2$ is given in Appendix and $\sigma_{\epsilon}^2 = \mathbb{E}(\epsilon_i^2\mid \mathbf{M}_{i\cdot})$ .
\label{prop: limiting distribution}
\end{Proposition}

\section{Invalid TCPs and OCPs}
\label{sec: invalid w}
We now extend to a scenario where, alongside a point exposure $D$, outcome $Y$, unmeasured confounder $U$, and $p_z$ candidate TCPs, we also have $p_w$ candidate OCPs, denoted $W_1, \ldots, W_{p_w}$. Some of these OCPs may be `invalid' in that 
they may not be independent of either the TCPs or the treatment itself, given the unmeasured confounder.

\vspace{2 mm}
\noindent 
As with TCPs, we assume that more than $50\%$ of the candidate OCPs are valid. Both assumptions are required for identification of the causal effect in our framework. That said, the estimation strategies differ: for TCPs, we rely on selection of a valid subset based on an identified moment condition; for OCPs, where the valid proxy is unknown, the majority condition allows us to combine estimators across candidates in a way that mitigates the influence of invalid proxies and ensures consistent estimation. 

\begin{table}[H]
    \begin{tabular}{|l|}
    \hline
    {\bf Step 1.} For each $j = 1, \ldots, p_w$, apply the algorithm in Table 1 using $W_j$ as the valid OCP to obtain the \\
    \hspace{44 pt} corresponding estimator $\widehat{\beta}^j_{post}$.\\
    \\
    {\bf Step 2.} After obtaining $p_w$ estimators $\widehat{\beta}^1_{post}, \widehat{\beta}^2_{post}, \ldots, \widehat{\beta}^{p_w}_{post}$, compute the final estimator as the median of \\
    \hspace{44 pt} these $p_w$ estimators:\\
    \\
    \hspace{195 pt} $\widehat{\beta}^{(p_w)} = \text{median}(\widehat{\beta}^1_{post}, \widehat{\beta}^2_{post}, \ldots, \widehat{\beta}^{p_w}_{post}).$\\
    \\
    \hline
    {\bf Note:} For each $j \in \{1, \ldots, p_w\}$, any quantity $T^j$ in Section \ref{sec: invalid w} is equivalent to the quantity $T$ in Section \ref{sec: invalid z},\\
    \hspace{35 pt} but corresponds specifically to $W_j$ instead of $W$. \\
    \hline
    \end{tabular}
    \caption{Estimation Algorithm in Presence of Invalid $Z$ and $W$}
    \label{tab: algo inv ZW}
\end{table}

\noindent The estimation procedure systematically evaluates each candidate OCP $W_j$, for $j = 1, \ldots, p_w$, by applying the algorithm described in Table \ref{tab: algo inv Z}. This involves selecting valid TCPs, using the identified invalid TCPs as measured confounders, and performing proximal 2SLS to estimate the causal effect. The final estimate is then obtained by taking the median of the resulting estimates across all $W_j$, leveraging the assumed majority validity of the OCPs to ensure robustness and accuracy.

\vspace{2 mm}
\noindent Based on the results in Section \ref{sec: invalid z}, if $W_j$  is valid, the resulting estimate of the causal effect will also be valid, exhibiting desirable properties such as consistency, asymptotic normality. Conversely, if $W_j$ is not valid, the estimator $\widehat{\beta}^j_{post}$ is expected to be biased, leading to an inaccurate estimation of the true causal effect $\beta$. In fact when $W_j$ is invalid, the resulting estimator $\widehat{\beta}^j_{post}$ is biased due to the unaccounted influence of hidden confounders and inconsistent TCP selection. Specifically, these biases can be characterized by constants $c_{1j}$ and $c_{2j}$, reflecting the limiting contributions of unmeasured confounding and invalid TCP selection, respectively (see Proof of Theorem \ref{thm: lim dist median est} for details).

\begin{Theorem}
Under assumptions \ref{ass a} and \ref{ass b}, the majority validity conditions for both TCPs and OCPs, and provided that the conditions of Proposition \ref{prop: limiting distribution} hold when the corresponding OCP under consideration is valid,
\begin{equation}
    \sqrt{n}\left(\widehat{\beta}^{(p_w)} - \beta\right) \to_d q_{[l],p_w-s_w}
\end{equation}
where $s_w$ is the number of invalid OCPs. If $p_w$ is odd, $q_{[l],p_w-s_w}$ is the $l-$th order statistic of the limiting normal distribution of $\sqrt{n}\left(\{\widehat{\beta}^j_{post}\}_{j=s_w +1}^{p_w} - \beta\right)$, where $l$ is determined by $p_w, s_w$ and the signs of $\{c_{1,j}, c_{2,j}\}_{1\leq j \leq p_w}$. If $p_w$ is even, $q_{[l], p_w-s_w}$ is defined as the average of either the $[l]$ and $[l-1]$-order statistics, or the $[l]$ and $[l+1]$-order statistics.
\label{thm: lim dist median est}
\end{Theorem}
Theorem \ref{thm: lim dist median est} establishes the asymptotic distribution of the final estimator $\widehat{\beta}^{(p_w)}$, constructed as the median of the candidate estimators $\{\widehat{\beta}^j_{post}\}$. The result highlights the robustness of the median-based aggregation strategy in the presence of invalid OCPs, provided that the majority validity assumption holds. Specifically, as long as more than $50\%$ of the candidate OCPs are valid, the estimator converges in distribution to an order statistic of the limiting normal distribution derived from the valid OCPs. Importantly, this result also guarantees that $\widehat{\beta}^{(p_w)}$ is $\sqrt{n}$-consistent for the true causal effect $\beta$, meaning that the estimation error shrinks at the parametric rate of $1/\sqrt{n}$. 

\vspace{3 mm}
\noindent \underline{\bf Inference via Subsampling:}
The estimator $\widehat{\beta}^{(p_w)}$, introduced in Table \ref{tab: algo inv ZW}, is constructed as the median of mutually dependent estimators, each corresponding to a different candidate OCP. This dependence structure makes standard asymptotic inference difficult. To address this, we adopt a nonparametric inference approach based on subsampling.

\vspace{2 mm}

\noindent In the subsampling framework (Politis, Romano, and Wolf, 1999), we repeatedly draw random subsets of the full dataset, each of size $b < n$, where $n$ is the total sample size. For each subset, the estimator $\widehat{\beta}^{(p_w)}$ is recalculated following the same steps outlined in Table \ref{tab: algo inv ZW}. Let $\widehat{\beta}^{(p_w)}_{(1)}, \widehat{\beta}^{(p_w)}_{(2)}, \ldots, \widehat{\beta}^{(p_w)}_{(N)}$ denote the resulting estimates from $N$ such subsamples. The empirical distribution of these subsample estimates serves as an approximation to the sampling distribution of $\widehat{\beta}^{(p_w)}$.

\vspace{2 mm}

\noindent A two-sided $100(1-\alpha)\%$ confidence interval for $\beta$ is then constructed as:

$$
\left[q^{*}_{\alpha/2},\; q^{*}_{1-\alpha/2}\right],
$$
where $q^{*}_{\alpha/2}$ and $q^{*}_{1-\alpha/2}$ are the empirical $\alpha/2$ and $1-\alpha/2$ quantiles of the subsample estimates $\{\widehat{\beta}^{(p_w)}_{(1)}, \ldots, \widehat{\beta}^{(p_w)}_{(N)}\}$.

\vspace{2 mm}

\noindent The validity of this subsampling-based confidence interval relies on mild conditions established in the general subsampling literature. Following the setup in Politis, Romano, and Wolf (1999), let $X_1,\dots,X_n$ be i.i.d. observations, and $\widehat{\theta}_n(X_1,\dots,X_n)$ an estimator of a parameter $\theta(P)$, where $P$ denotes the underlying data-generating distribution. The required assumption is:

\vspace{2 mm}

\noindent {\bf (A1)} Define

$$
J_n(x, P) := \mathbb{P}_P\left[\tau_n\left(\widehat{\theta}_n(X_1,\ldots,X_n) - \theta(P)\right) \leq x\right].
$$
There exists a non-degenerate limiting distribution $J(P)$ such that $J_n(P) \rightsquigarrow J(P)$ as $n \to \infty$.

\vspace{2 mm}

\noindent No further assumptions on $\widehat{\theta}_n$ are required. In our setting, $\widehat{\theta}_n$ corresponds to $\widehat{\beta}^{(p_w)}$, with the scaling $\tau_n = \sqrt{n}$. Proposition \ref{prop: limiting distribution} confirms that this assumption holds for our estimator. Therefore, the subsampling procedure provides valid large-sample inference for $\beta$, despite the estimator’s non-regular structure. The remaining step is to appropriately choose the number of subsamples $ N $ and the subsample size $ b $. As a general guideline, we set $ N = 1000 $ and $ b = n^{4/5} $.

\section{Simulation Study}

\subsection{Invalid TCPs}

This section evaluates the estimator in \eqref{eq: post-ad LASSO} when a valid OCP is present and the set of TCPs includes both valid and invalid proxies, under a majority-rule setting where most TCPs are valid. Here, we consider $p_z = 10$ candidate TCPs, with the first $s_z = 3$ being invalid. The sample size $n$ varies as $n \in \{1500, 2500, 5000, 10000\}$. Specifically, for $1 \leq i \leq n$ and $1 \leq j \leq p_z$, we use the following data-generating process:

\begin{equation}
    \begin{aligned}
        U_i & \sim_{i.i.d.} N(0, \frac{1}{2}) \\
        Z_{ij} & = 0.25 + U_i + \epsilon^z_{ij} \text{ where } \epsilon^z_{ij} \sim_{i.i.d.} N(0, \frac{1}{2}) \\
        W_i & = 0.25 + U_i + \epsilon^w_i \text{ where } \epsilon^w_i \sim_{i.i.d.} N(0, \frac{1}{2}) \\
        D_i & = 0.25 + 0.2 U_i + \mathbf{Z}_{i\cdot}^{\intercal}\boldsymbol \xi^z + \epsilon_i^d \text{ where } \epsilon_i^d \sim_{i.i.d.} N(0,1) \\
        Y_i & = 0.25 + D_i\beta + 0.2U_i + \mathbf{Z}_{i\cdot}^{\intercal}\boldsymbol \alpha + \epsilon_i^y \text{ where } \epsilon_i^y \sim_{i.i.d.} N(0,1)
    \end{aligned}
    \label{eq: DGP Z}
\end{equation}
Here, following the IV nomenclature, the strength parameter $\xi^z$ is set as
\begin{equation*}
    \xi^z_j =
    \begin{cases}
        0.6 & \text{ if } 1 \leq j \leq s_z \\
        0.2 & \text{ if } s_z + 1 \leq j \leq p_z,
    \end{cases}
\end{equation*}
and the invalidity of the candidate TCPs are reflected by $\alpha$.
\begin{equation*}
    \alpha_j =
    \begin{cases}
        0.8 & \text{ if } 1 \leq j \leq s_z \\
        0 & \text{ if } s_z + 1 \leq j \leq p_z.
    \end{cases}
\end{equation*}
The error terms $\epsilon^z, \epsilon^w, \epsilon^d, \epsilon^y$ are assumed mutually independent.

\vspace{2 mm}
\noindent Table \ref{tab: Inv Z 3} reports the estimation results and the coverage probabilities of the confidence intervals constructed using \eqref{eq: CI Z}. To benchmark the performance of our proposed estimator, we compare it against three alternative methods:
\begin{itemize}
\item \textbf{Oracle 2SLS:} Assumes perfect knowledge of which TCPs are invalid and correctly adjusts for them as confounders, as defined in \eqref{eq: oracle Z}.
\item \textbf{Unadjusted 2SLS:} Ignores potential invalidity and treats all candidate TCPs as valid.
\item \textbf{Ordinary Least Squares (OLS):} Directly regresses the outcome on the treatment without accounting for unmeasured confounding.
\end{itemize}

\begin{table}[htp!]
\centering
\scalebox{0.65}{
\begin{tabular}{|r|rrrrr|rrrrr|rrrrr|rrrrr|}
\hline
 &\multicolumn{5}{c|}{Adaptive Proximal} &\multicolumn{5}{c|}{Oracle} &\multicolumn{5}{c|}{U2SLS} &\multicolumn{5}{c|}{OLS} \\
\hline
$n$& Cov & Len & Bias & SE & RMSE & Cov & Len & Bias & SE & RMSE & Cov & Len & Bias & SE & RMSE & Cov & Len & Bias & SE & RMSE \\
 \hline
 1500 & 0.93 & 0.016 & 0.003 & 0.020 & 0.020 & 0.94 & 0.010 & 0.000 & 0.003 & 0.003 & 0.04 & 0.127 & 0.162 & 0.042 & 0.162 & 0.00 & 0.038 & 0.607 & 0.009 & 0.607 \\
 2500 & 0.94 & 0.009 & 0.000 & 0.004 & 0.004 & 0.94 & 0.008 & 0.000 & 0.002 & 0.002 & 0.00 & 0.099 & 0.156 & 0.031 & 0.156 & 0.00 & 0.030 & 0.607 & 0.007 & 0.607 \\
 5000 & 0.95 & 0.005 & 0.000 & 0.001 & 0.001 & 0.94 & 0.005 & 0.000 & 0.001 & 0.001 & 0.00 & 0.070 & 0.152 & 0.023 & 0.152 & 0.00 & 0.021 & 0.607 & 0.005 & 0.607 \\
 \hline
\end{tabular}
}
\caption{Results from $p_z = 10$ candidate TCPs, with the first $s_z = 3$ being invalid. The columns indexed with ``Cov" and ``Len" represent the empirical coverage and length of the CIs; the columns indexed with ``RMSE", ``Bias" and ``SE" represent the RMSE, bias and standard error, respectively. The columns under ``Adaptive Proximal" ,``Oracle", ``U2SLS" and ``OLS" correspond to the proposed estimator, the oracle 2SLS, the unadjusted 2SLS and OLS estimators respectively.}
\label{tab: Inv Z 3}
\end{table}

We can conclude from Table \ref{tab: Inv Z 3} that both the proposed and oracle estimators maintain well-calibrated confidence intervals with empirical coverage close to the nominal level, negligible bias, and low RMSE, with performance improving as sample size increases. The proposed estimator has slightly longer confidence intervals than the oracle, reflecting its lack of access to oracle information, but this difference diminishes with larger samples. In contrast, the unadjusted 2SLS estimator performs poorly, exhibiting severe under-coverage, substantial bias, and high RMSE. This is because it incorrectly treats all TCPs as valid, thereby ignoring the confounding role of the invalid TCPs. As a result, these variables are omitted from the outcome regression in the second stage of 2SLS, leading to omitted variable bias. The OLS estimator shows the lowest performance, with zero coverage and substantial omitted variable bias, as it overlooks the presence of confounders. This poor performance highlights the problem of unmeasured confounders, which proximal causal inference aims to address. Overall, the proposed estimator emerges as a robust and practical alternative to the oracle, achieving competitive efficiency without oracle knowledge.

\begin{table}[htp!]
\centering
\scalebox{0.7}{
\begin{tabular}{|r|rrrrr|rrrrr|rrrrr|}
\hline
 &\multicolumn{5}{c|}{Adaptive Proximal} &\multicolumn{5}{c|}{Oracle} &\multicolumn{5}{c|}{U2SLS} \\
\hline
$s_z$& Cov & Len & Bias & SE & RMSE & Cov & Len & Bias & SE & RMSE & Cov & Len & Bias & SE & RMSE  \\
 \hline
 1 & 0.94 & 0.008 & 0.001 & 0.006 & 0.006 & 0.93 & 0.008 & 0.000 & 0.002 & 0.002 & 0.00 & 0.046 & 0.070 & 0.016 & 0.072 \\
 2 & 0.93 & 0.008 & 0.000 & 0.002 & 0.002 & 0.92 & 0.008 & 0.008 & 0.002 & 0.002 & 0.00 & 0.072 & 0.121 & 0.024 & 0.124 \\
 3 & 0.94 & 0.009 & 0.000 & 0.004 & 0.004 & 0.94 & 0.008 & 0.000 & 0.002 & 0.002 & 0.00 & 0.099 & 0.156 & 0.031 & 0.156 \\
 4 & 0.94 & 0.009 & 0.001 & 0.009 & 0.009 & 0.95 & 0.008 & 0.000 & 0.002 & 0.002 & 0.02 & 0.127 & 0.171 & 0.040 & 0.171 \\
 5 & 0.93 & 0.281 & -0.047 & 0.072 & 0.086 & 0.94 & 0.008 & 0.000 & 0.002 & 0.002 & 0.04 & 0.154 & 0.183 & 0.050 & 0.183 \\
 6 & 0.95 & 0.352 & -0.021 & 0.020 & 0.090 & 0.95 & 0.008 & 0.000 & 0.002 & 0.002 & 0.12 & 0.188 & 0.175 & 0.057 & 0.175 \\
 7 & 0.95 & 0.368 & -0.019 & 0.095 & 0.097 & 0.95 & 0.008 & 0.000 & 0.002 & 0.002 & 0.24 & 0.222 & 0.161 & 0.062 & 0.172 \\
 8 & 0.95 & 0.374 & -0.016 & 0.101 & 0.103 & 0.96 & 0.008 & 0.000 & 0.002 & 0.002 & 0.47 & 0.261 & 0.131 & 0.070 & 0.149 \\
 \hline
\end{tabular}
}
\caption{Results from $p_z = 10$ candidate TCPs, with an increasing number of invalid proxies.  Sample size is fixed at $n=2500$. The columns indexed with ``Cov" and ``Len" represent the empirical coverage and length of the CIs; the columns indexed with ``RMSE", ``Bias" and ``SE" represent the RMSE, bias and standard error, respectively; the column under ``$s_z$" represents the number of invalid TCPs. The columns under ``Adaptive Proximal", ``Oracle", and ``U2SLS" correspond to the proposed estimator, the oracle and the unadjusted 2SLS estimators respectively.}
\label{tab: Inv Z sz}
\end{table}

\noindent Table \ref{tab: Inv Z sz} evaluates the performance of the proposed, oracle, and unadjusted 2SLS estimators as the number of invalidTCPs ($s_z$) varies across $\{1, 2, \ldots, 8\}$, with $n = 2500$. Note that as long as $s_z < 5$, the majority validity assumption for TCP holds. The findings align with those from Table \ref{tab: Inv Z 3}. The proposed estimator maintains empirical coverage close to the nominal level and competitive RMSE, demonstrating robustness to increasing $s_z$, although its confidence interval length and standard error grow slightly as $s_z$ increases, reflecting added uncertainty. The oracle estimator consistently outperforms the others, achieving the shortest intervals, negligible bias, low RMSE, and near-perfect coverage due to its access to true information about invalid TCPs. In contrast, the unadjusted estimator suffers from severe omitted variable bias in the second stage of 2SLS, leading to substantial bias, poor coverage (as low as 0.02), wide intervals, and high RMSE, particularly at higher $s$. Overall, the proposed estimator proves to be a practical and reliable alternative to the oracle, effectively addressing the challenges posed by invalid TCPs, while the unadjusted estimator remains unsuitable in such scenarios.

\subsection{Invalid TCPs and OCPs}
\label{sec: Simu Inv WZ}

In this section, we extend the data-generating process in \eqref{eq: DGP Z} by generating $p_w = 10$ candidate OCPs, denoted $W_1, \ldots, W_{10}$, instead of a single valid OCP. Among these, the first $s_w = 3$ proxies are set to be invalid in the sense that they directly depend on the treatment $D$, which is encoded through the corresponding non-zero entries of $\boldsymbol{\xi}^w$. Specifically, for $1 \leq k \leq p_w = 10$, we generate:

$$
W_{ik} = 0.25 + U_i + A_i \xi^w_k + \epsilon^w_{ik}, \quad \epsilon^w_{ik} \overset{i.i.d.}{\sim} N\left(0,\frac{1}{2}\right),
$$

where

$$
\xi^w_k =
\begin{cases}
0.8 & \text{if } 1 \leq k \leq s_w, \\
0 & \text{if } s_w + 1 \leq k \leq p_w.
\end{cases}
$$
This setup ensures that only the first three candidate OCPs are invalid by construction, as their values depend on the treatment $D$, while the remaining seven are valid OCPs independent of $D$.

\vspace{2 mm}
\begin{table}[htp!]
\centering
\scalebox{0.65}{
\begin{tabular}{|r|rrrrr|rrrrr|rrrrr|rrrrr|}
\hline
 &\multicolumn{5}{c|}{Adaptive Proximal} &\multicolumn{5}{c|}{Oracle} &\multicolumn{5}{c|}{U2SLS} & \multicolumn{5}{c|}{OLS} \\
\hline
$n$& Cov & Len & Bias & SE & RMSE & Cov & Len & Bias & SE & RMSE & Cov & Len & Bias & SE & RMSE & Cov & Len & Bias & SE & RMSE  \\
 \hline
 1500 & 0.96 & 0.073 & 0.031 & 0.044 & 0.054 & 1 & 0.075 & 0.006 & 0.091 & 0.091 & 1 & 1182.35 & 41 & 749.53 & 750.65 & 0 & 0.038 & 0.608 & 0.009 & 0.608 \\
 2500 & 0.99 & 0.038 & 0.014 & 0.032 & 0.035 & 1 & 0.063 & 0.003 & 0.054 & 0.054 & 1 & 708.14 & 32 & 714.14 & 714.89 & 0 & 0.029 & 0.608 & 0.008 & 0.608 \\
 5000 & 0.97 & 0.012 & 0.002 & 0.014 & 0.014 & 1 & 0.033 & 0.001 & 0.041 & 0.041 & 1 & 586.23 & 19 & 337.03 & 337.59 & 0 & 0.021 & 0.607 & 0.005 & 0.607 \\
 \hline
\end{tabular}
}
\caption{Results from $p_z = 10$ candidate TCPs and $p_w=10$ candidate OCPs, with the first $s_z = 3$ and $s_w=3$ being invalid. The columns indexed with ``Cov" and ``Len" represent the empirical coverage and length of the CIs; the columns indexed with ``RMSE", ``Bias" and ``SE" represent the RMSE, bias and standard error, respectively. The columns under ``Adaptive Proximal" ,``Oracle", ``U2SLS" and ``OLS" correspond to the proposed estimator, the oracle, the unadjusted 2SLS and the OLS estimators respectively.}
\label{tab: Inv WZ other}
\end{table}

\noindent Table \ref{tab: Inv WZ other} compares the performance of the proposed, oracle, unadjusted 2SLS, and OLS estimators. The confidence intervals, constructed using the subsampling procedure, exhibit valid but slightly conservative coverage, with empirical coverage exceeding the nominal level. This over-coverage may be attributed to the choice of subsampling hyperparameters, such as subsample size and the number of subsamples. Bias, standard error, RMSE, and interval length decrease as sample size ($n$) increases, indicating improved accuracy and precision.The oracle estimator, with access to true information about invalid proxies, consistently achieves the lowest bias and RMSE. In contrast, the unadjusted estimator performs poorly, suffering from extreme bias and RMSE due to its failure to address invalid proxies. It yields abnormally wide CIs, with extremely large lengths (e.g., over 1000 at $n=1500$), leading to artificially perfect coverage. These intervals are practically uninformative and reflect the estimator's instability in the presence of invalid proxies.  The OLS estimator also exhibits high RMSE driven by substantial bias. It produces narrow intervals with zero empirical coverage, revealing its complete failure to provide valid inference under endogeneity.
Overall, the results reinforce the effectiveness of the proposed estimator in managing invalid proxies.

\begin{table}[htp!]
\centering
\scalebox{0.7}{
\begin{tabular}{|r|r|rrr|}
\hline
 &&\multicolumn{3}{c|}{Adaptive Proximal} \\
\hline
 $s_z$ & $s_w$ & Bias & SE & RMSE \\
 \hline
 \multirow{4}{*}{3} & 3 & 0.014 & 0.032 & 0.035\\
  & 4 & 0.003 & 0.024 & 0.024 \\
  & 5 & -0.111 & 0.099 & 0.149 \\
  & 6 & -0.233 & 0.218 & 0.319 \\
 \hline
 \multirow{4}{*}{4} & 3 & 0.034 & 0.051 & 0.061 \\
  & 4 & 0.005 & 0.047 & 0.047 \\
  & 5 & -0.207 & 0.184 & 0.278 \\
  & 6 & -0.444 & 0.416 & 0.609 \\
 \hline
 \multirow{4}{*}{5} & 3 & 0.095 & 0.049 & 0.107 \\
 & 4 & 0.041 & 0.047 & 0.062 \\
 & 5 & -0.598 & 0.139 & 0.614 \\
 & 6 & -1.302 & 0.324 & 1.341 \\
 \hline
 \multirow{4}{*}{6} & 3 & 0.106 & 0.052 & 0.118 \\
 & 4 & 0.016 & 0.062 & 0.065 \\
 & 5 & -0.981 & 0.178 & 0.996 \\
 & 6 & -2.270 & 0.292 & 2.289 \\
 \hline
\end{tabular}
}
\caption{Results from $p_z = 10$ candidate TCPs and $p_w=10$ candidate OCPs, with an increasing number of invalid proxies. Sample size is fixed at $n=2500$. The columns indexed with ``Cov" and ``Len" represent the empirical coverage and length of the CIs; the columns indexed with ``RMSE", ``Bias" and ``SE" represent the RMSE, bias and standard error, respectively; the columns under ``$s_z$" and ``$s_w$" represent the number of invalid TCPs and OCPs respectively.. The column under ``Adaptive Proximal" corresponds to our proposed estimator.}
\label{tab: Inv WZ sz}
\end{table}

\vspace{2 mm}
\noindent Table \ref{tab: Inv WZ sz} analyzes the performance of the proposed estimator under varying numbers of invalid TCPs ($s_z$) and invalid OCPs ($s_w$) with sample size fixed at $n = 2500$. The bias, SE, and RMSE generally increase as $s_z$ or $s_w$ grows, reflecting the challenges posed by an increasing number of invalid proxies. For smaller values of $s_z$ and $s_w$ (e.g., $s = 3, s_w \in \{3, 4\}$), the estimator maintains low bias and RMSE, demonstrating robustness. However, as both $s_z$ and $s_w$ increase beyond these thresholds, bias dominates the RMSE, particularly for larger $s_w$. For example, at $s_z = 5, s_w = 6$, the bias (-1.302) and RMSE (1.341) indicate a significant deterioration in performance. These results highlight the estimator’s sensitivity to the number of invalid proxies and the importance of adhering to majority rule for reliable performance.

\section{Real Data Application}

In this section, we investigate the potential impact of right heart catheterization (RHC) on 30-day survival among critically ill patients in the ICU, using data from the SUPPORT study (Study to Understand Prognoses and Preferences for Outcomes and Risks of Treatments). RHC is an invasive procedure in which a thin tube (catheter) is inserted into the right side of the heart and the pulmonary artery to obtain detailed measurements of heart and lung function, which can help guide treatment in critically ill patients. The analysis primarily focuses on determining whether undergoing RHC within 24 hours of ICU admission influences patient survival, with 2184 patients receiving RHC and 3551 who did not undergo the procedure. This dataset has been previously analyzed in \cite{introeric}, where potential confounding has been adjusted by leveraging a rich set of 73 covariates. Notably, 10 physiological markers—such as serum sodium, creatinine, and hematocrit—were collected during the first 24 hours of ICU admission. These markers are related to the patient's overall physiological condition and are believed to serve as proxies for the unmeasured underlying physiological status, which is an important confounder in this context. Among the 10 markers, four measures—pafi1, paco21, ph1, and hema1—exhibit strong associations with both the treatment and the outcome, making them particularly relevant for constructing the proxy variables, TCPs and OCPs. The remaining six physiological markers, alongside additional clinical and demographic variables, are included as covariates (X) to further adjust for confounding. \cite{introeric} proposes a structured method to classify candidate proxy variables into TCP and OCP categories based on their associations with the treatment and the outcome, using regression-based rankings to guide the allocation. This is followed by the application of the P2SLS algorithm to estimate the causal effect of RHC.

\vspace{2 mm}
\noindent 
Instead, we examine 10 distinct configurations, each treating one of the physiological proxies as the valid OCP ($W$) and the remaining nine as candidate TCPs ($\mathbf{Z}$). For instance, when $W = {\text{ph1}}$, the candidate TCPs are ${\text{pafi1}, \text{paco21}, \text{hema1}, \text{sod1}, \text{pot1}, \text{crea1}, \text{bili1}, \text{alb1}, \text{wblc1}}$. We then apply our procedure from Table \ref{tab: algo inv Z} to test the validity of the candidate TCPs under the assumption that $W$ is a valid OCP. This is repeated across all 10 assignments, ensuring that all proxies are considered in both roles. The results are reported in Table \ref{tab: real inv z}.

\begin{table}[ht!]
\centering
\begin{tabular}{|c|c|c|c|c|}
\hline
W & Invalid TCPs & Valid TCPs & $\widehat{\beta}$ & CI \\ \hline
ph1 & \{bili1\} & \{hema1, pafi1, paco21, sod1, pot1, crea1, alb1, wblc1\} & -1.405 & $[-1.935,-0.876]$ \\ \hline
hema1 & \{ph1, bili1\} & \{hema1, pafi1, paco21, sod1, pot1, crea1, alb1, wblc1\} & -1.378 & $[-1.913,-0.844]$ \\ \hline
pafi1 & \{ph1, bili1\} & \{hema1, pafi1, paco21, sod1, pot1, crea1, alb1, wblc1\} & -1.809 & $[-2.503,-1.115]$ \\ \hline
paco21 & \{ph1, bili1\} & \{hema1, pafi1, paco21, sod1, pot1, crea1, alb1, wblc1\} & -1.568 & $[-2.135,-1.000]$ \\ \hline
sod1 & \{ph1, bili1\} & \{hema1, pafi1, paco21, sod1, pot1, crea1, alb1, wblc1\} & -1.376 & $[-1.917,-0.834]$ \\ \hline
pot1 & \{crea1, bili1\} & \{hema1, pafi1, paco21, sod1, pot1, bili1, alb1, wblc1\} & -1.509 & $[-2.073,-0.946]$ \\ \hline
crea1 & \{ph1, bili1\} & \{hema1, pafi1, paco21, sod1, pot1, crea1, alb1, wblc1\} & -1.399 & $[-1.928,-0.870]$ \\ \hline
bili1 & \{ph1\} & \{hema1, pafi1, paco21, sod1, pot1, crea1, alb1, wblc1\} & -1.332 & $[-1.892,-0.772]$ \\ \hline
alb1 & \{ph1, bili1\} & \{hema1, pafi1, paco21, sod1, pot1, crea1, alb1, wblc1\} & -1.430 & $[-1.964,-0.896]$ \\ \hline
wblc1 & \{ph1, bili1\} & \{hema1, pafi1, paco21, sod1, pot1, crea1, alb1, wblc1\} & -1.310 & $[-1.872,-0.747]$ \\ \hline
\end{tabular}
\caption{Summary of estimates and confidence intervals. Here, $W$ represents the chosen valid outcome-inducing proxy.
``Invalid TCPs" and ``Valid TCPs" identify the estimated sets of invalid and valid TCPs respectively, $\widehat{\beta}$ provides the estimated causal effect, and CI indicates the corresponding confidence interval.}
\label{tab: real inv z}
\end{table}

\vspace{3 mm}
\noindent From Table \ref{tab: real inv z}, it can be concluded that under the majority rule framework, where at most $4$ out of $9$ candidate proxies can be invalid, $\text{bili1}$ and $\text{ph1}$ consistently rank as invalid TCPs. 
Interestingly, most of the remaining markers are never flagged as invalid under any configuration, suggesting greater reliability. These results are consistent with clinical understanding. For example, bilirubin ($\text{bili1}$) reflects liver function and is thus a good indicator of a patient’s physiological state; however, elevated bilirubin levels can directly affect survival through complications like jaundice, making it an invalid TCP. In contrast, markers like white blood cell count ($\text{wblc1}$), which indicate the presence of infection, are not known to directly affect mortality, making them more plausible candidates for valid TCPs.

\vspace{2 mm}
\noindent Across the ten configurations, the estimated causal effect $\widehat{\beta}$ ranges from –1.809 (with $W = \text{pafi1}$) to –1.310 (with $W = \text{wblc1}$), with all corresponding confidence intervals excluding zero. This indicates a statistically significant negative effect of RHC on 30-day survival. These findings are consistent with prior evidence suggesting that RHC, as an invasive intervention, may adversely impact survival among critically ill patients.

\vspace{5 mm}
\noindent After generating the 10 estimates of the causal effect, each corresponding to one marker treated as OCP and the remaining 9 as candidate TCPs, we plan to take the median of these 10 estimates to ensure a robust and consistent causal effect estimator. For this approach to be valid, we need the assumption that more than $50\%$ of the candidate proxies are valid OCPs. However, a conceptual conflict arises: by definition, a valid TCP cannot directly affect the outcome ($Y$), though it may be associated with the treatment ($D$), while a valid OCP cannot be directly related to the treatment, although it may affect the outcome. This asymmetry implies that a proxy cannot simultaneously satisfy the criteria for both a valid TCP and a valid OCP, seemingly ruling out the possibility of overlap. To resolve this, we introduce the \emph{Disconnected Proxy Assumption (DPA)}, under which a valid proxy—whether TCP or OCP—is assumed to have no direct effect on either the treatment or the outcome. This assumption corresponds to a special case of the \emph{No-Connection (NC)} condition validated in the DANCE framework (\cite{DANCE}), where the graphical criterion ensures that the selected proxies do not exert direct influence on the treatment or the outcome. We refer to this assumption as DPA to highlight its role in disconnecting the valid proxies from the core causal pathway.
This broader assumption allows overlap, where a single proxy can be both a valid TCP and OCP. As a result, we can consistently assume that more than $50\%$ of the proxies are valid without conflict, justifying the use of the median estimator for causal effect estimation.


\vspace{5 mm}
\noindent Thus, under the additional assumption that at least five of the proxies $\text{ph1}, \text{pafi1}, \text{paco21}, \text{hema1}, \text{sod1}, \text{pot1},$ $ \text{crea1}, \text{bili1}, \text{alb1}, \text{wblc1}$ are valid OCPs, we apply the aggregation technique described in Section \ref{sec: invalid w} by taking the median of the 10 causal effect estimates from the previous cases. This yields an estimated causal effect of $-1.402$, with a subsampling-based confidence interval of $[-2.628, -0.218]$. Since the interval excludes zero, the effect is statistically significant, providing evidence that RHC has a significantly negative impact on 30-day survival.

\vspace{2 mm}
\noindent For comparison, \cite{semiparam} report two influence function-based doubly robust estimates using the same SUPPORT dataset. The first, which relies on the standard exchangeability assumption, yields an estimated causal effect of $-1.17$ with a $95\%$ confidence interval of $[-1.79, -0.55]$. The second, based on a proximal causal inference approach incorporating bridge functions and proxy variables, estimates the effect to be $-1.66$ with a $95\%$ confidence interval of $[-2.50, -0.83]$. In contrast, our estimate is $-1.402$, with a subsampling-based confidence interval of $[-2.628, -0.218]$, falling between the two. While Cui et al. identify proxies through regression-based relevance and pre-assign roles for TCPs and OCPs, we consider all 10 physiological markers simultaneously. Our approach assumes a majority rule for both TCP and OCP identification and permits overlap between valid TCPs and OCPs under a ``disconnected proxies" assumption. This aggregation-based strategy provides a flexible and interpretable alternative that avoids strong structural assumptions, while still yielding statistically significant evidence of a negative causal effect of RHC on 30-day survival.

\section{Conclusion}

This paper introduced a robust framework for proximal causal inference when the validity of proxy variables is uncertain. By leveraging penalized estimation techniques and insights from the instrumental variable literature, we developed methods to identify and select valid proxies, ensuring reliable causal effect estimation even in the presence of invalid ones. Our theoretical results established the consistency and asymptotic properties of the proposed estimators, and their practical utility was demonstrated through an application to the SUPPORT study. The findings highlight the flexibility of our approach in handling real-world confounding complexities. 

\vspace{5 mm}
\noindent Looking ahead, one important direction for future research is to extend our framework beyond the linear structural equation model. In particular, developing methods that accommodate non-linear models—or ideally, a fully non-parametric setup—would significantly broaden the applicability of our approach across a wider range of real-world scenarios.

\appendix
\appendix

\section{Asymptotic Variance of the Oracle Estimator}
\label{app:oracle_variance}

This appendix provides the asymptotic variance expression for the oracle estimator $\widehat{\beta}_{or}$ introduced in equation~\eqref{eq: oracle Z}.

\subsection{Asymptotic Distribution}

Under standard regularity conditions, the oracle estimator satisfies the following asymptotic normality:
\[
\sqrt{n}(\widehat{\beta}_{or} - \beta) \xrightarrow{d} \mathcal{N}(0, \sigma^2_{or}),
\]
where the asymptotic variance $\sigma^2_{or}$ is given by:
\begin{small}
\begin{equation}
\begin{aligned}
    \sigma^2_{or} = \sigma^2_{\epsilon} \cdot \bigg[ & \E(D_i^2) - \E\left(D_i\widehat{\mathbf{N}}_{A_{i\cdot}}^{\intercal}\right)
    \left\{\E\left(\widehat{\mathbf{N}}_{A_{i\cdot}}\widehat{\mathbf{N}}_{A_{i\cdot}}^{\intercal}\right)\right\}^{-1}
    \E\left(\widehat{\mathbf{N}}_{A_{i\cdot}}\mathbf{M}_{i\cdot}^{\intercal}\right)
    \left\{\E\left(\mathbf{M}_{i\cdot}\mathbf{M}_{i\cdot}^{\intercal}\right)\right\}^{-1} \\
    & \cdot \E\left(\mathbf{M}_{i\cdot}\widehat{\mathbf{N}}_{A_{i\cdot}}^{\intercal}\right)
    \left\{\E\left(\widehat{\mathbf{N}}_{A_{i\cdot}}\widehat{\mathbf{N}}_{A_{i\cdot}}^{\intercal}\right)\right\}^{-1}
    \E\left(\widehat{\mathbf{N}}_{A_{i\cdot}}D_i\right) \bigg]^{-2} \cdot V,
\end{aligned}
\end{equation}
\end{small}

where $V$ is the expression:
\begin{small}
\begin{equation}
\begin{aligned}
V = \E(D_i^2) & + \E\left(D_i\widehat{\mathbf{N}}_{A_{i\cdot}}^{\intercal}\right)
\left\{\E\left(\widehat{\mathbf{N}}_{A_{i\cdot}}\widehat{\mathbf{N}}_{A_{i\cdot}}^{\intercal}\right)\right\}^{-1}
\E\left(\widehat{\mathbf{N}}_{A_{i\cdot}}\mathbf{M}_{i\cdot}^{\intercal}\right)
\left\{\E\left(\mathbf{M}_{i\cdot}\mathbf{M}_{i\cdot}^{\intercal}\right)\right\}^{-1} \\
& \cdot \E\left(\mathbf{M}_{i\cdot}\widehat{\mathbf{N}}_{A_{i\cdot}}^{\intercal}\right)
\left\{\E\left(\widehat{\mathbf{N}}_{A_{i\cdot}}\widehat{\mathbf{N}}_{A_{i\cdot}}^{\intercal}\right)\right\}^{-1}
\E\left(\widehat{\mathbf{N}}_{A_{i\cdot}}D_i\right) \\
& - 2 \cdot \E\left(D_i\widehat{\mathbf{N}}_{A_{i\cdot}}^{\intercal}\right)
\left\{\E\left(\widehat{\mathbf{N}}_{A_{i\cdot}}N_{i\cdot}^{\intercal}\right)\right\}^{-1}
\E\left(\widehat{\mathbf{N}}_{A_{i\cdot}}\mathbf{M}_{i\cdot}^{\intercal}\right)
\left\{\E\left(\mathbf{M}_{i\cdot}\mathbf{M}_{i\cdot}^{\intercal}\right)\right\}^{-1}
\E\left(\mathbf{M}_{i\cdot}D_i\right).
\end{aligned}
\label{eq:oracle_variance}
\end{equation}
\end{small}

\subsection{Derivation Intuition}

This result follows from standard asymptotic arguments and projection-based decomposition in linear models. In particular, under the oracle model with the invalid TCPs correctly identified and moved to the regression, the estimator satisfies:
\begin{equation}
\begin{aligned}
\widehat{\beta}_{or} 
&= \frac{\mathbf{D}^{\intercal}\mathbf{P}_{\widehat{\mathbf{N}}_A^{\perp}}\mathbf{Y}}{\mathbf{D}^{\intercal}\mathbf{P}_{\widehat{\mathbf{N}}_A^{\perp}}\mathbf{D}} \\
&= \frac{\mathbf{D}^{\intercal}\mathbf{P}_{\widehat{\mathbf{N}}_A^{\perp}}\left(\mathbf{D}\beta + \mathbf{Z}_A\boldsymbol{\alpha}_A + \mathbf{W}\gamma + \boldsymbol{\epsilon}\right)}{\mathbf{D}^{\intercal}\mathbf{P}_{\widehat{\mathbf{N}}_A^{\perp}}\mathbf{D}} \\
&= \beta + \frac{\mathbf{D}^{\intercal}\mathbf{P}_{\widehat{\mathbf{N}}_A^{\perp}}\boldsymbol{\epsilon}}{\mathbf{D}^{\intercal}\mathbf{P}_{\widehat{\mathbf{N}}_A^{\perp}}\mathbf{D}},
\end{aligned}
\label{eq:oracle_steps}
\end{equation}
where $\boldsymbol{\epsilon}$ are i.i.d. with $\E[\epsilon_i \mid D_i, \mathbf{Z}_i] = 0$ and $\E[\epsilon_i^2 \mid D_i, \mathbf{Z}_i] = \sigma^2_{\epsilon}$. 

The final equality follows because $\mathbf{D}^{\intercal}\mathbf{P}_{\widehat{N}_A^{\perp}}\mathbf{W} = 0$ due to orthogonality:
\[
\mathbf{D}^{\intercal}\mathbf{P}_{\widehat{W}^{\perp}}\mathbf{W} 
= \mathbf{D}^{\intercal}\mathbf{W} - \mathbf{D}^{\intercal}\mathbf{P}_{\widehat{W}}\mathbf{W} 
= \mathbf{D}^{\intercal}\mathbf{W} - \mathbf{D}^{\intercal}\widehat{\mathbf{W}} = 0,
\]
since $\widehat{\mathbf{W}} = \mathbf{P}_M \mathbf{W}$ and $\mathbf{D}$ is in the column space of $\mathbf{M}$. The rest of the proof will follow by applying projection matrix algebra and the law of large numbers to the expression in \eqref{eq:oracle_steps}.

\section{Proofs}

\subsection{Proof of Corollary \ref{corr: majority}}

Suppose $I \leq \frac{p_z}{2}$. Then each set $C_m \subseteq \{1,\ldots,p_z\}$ considered in Theorem \ref{thm: identification} has size

$$
|C_m| = p_z - I + 1 \geq \frac{p_z}{2} + 1.
$$

By the pigeonhole principle, any two such sets $C_m$ and $C_{m'}$ must have a non-empty intersection, since otherwise their combined size would exceed $p_z$.

Let $j \in C_m \cap C_{m'}$. From the defining relation in Theorem \ref{thm: identification}, we have

$$
\widetilde{\boldsymbol{\delta}}_{j\cdot} q_m = \widetilde{\boldsymbol{\Gamma}}_j = \widetilde{\boldsymbol{\delta}}_{j\cdot} q_{m'},
$$

and since $\widetilde{\boldsymbol{\delta}}_{j\cdot} \neq 0$, it follows that $q_m = q_{m'}$.

Because this holds for any pair $(m, m')$, all $q_m$ are equal, ensuring the uniqueness of the solution to \eqref{eq: simple moment} by Theorem \ref{thm: identification}.

\subsection{Proof of Theorem \ref{thm: LASSO 2 step}}

Define $\widehat{W} := P_M W,$ where $M := (Z, D)$ and $P_A$ denotes the orthogonal projection onto the column space of matrix $A$. The parameters $\alpha, \beta, \gamma$ are estimated by solving

$$
(\widehat{\alpha}, \widehat{\beta}, \widehat{\gamma}) = \arg\min_{\alpha, \beta, \gamma} \frac{1}{2} \left\| P_M \big( Y - Z \alpha - D \beta - W \gamma \big) \right\|_2^2 + \lambda \|\alpha\|_1.
$$
Let $\gamma_1 := (\alpha^\intercal, \beta)^\intercal$ and note that $M \gamma_1 = Z \alpha + D \beta$. The objective becomes

$$
\frac{1}{2} \left\| P_M (Y - M \gamma_1 - W \gamma) \right\|_2^2 + \lambda \|\alpha\|_1.
$$
Using $\widehat{W} = P_M W$ and the property $P_{\widehat{W}} P_M = P_{\widehat{W}}$, we rewrite

$$
P_M (Y - M \gamma_1 - W \gamma) = P_M Y - M \gamma_1 - \widehat{W} \gamma.
$$
Decompose the norm into orthogonal components:

$$
\frac{1}{2} \left\| P_{\widehat{W}} (P_M Y - M \gamma_1) - P_{\widehat{W}} \widehat{W} \gamma \right\|_2^2 + \frac{1}{2} \left\| P_{\widehat{W}^\perp} (P_M Y - M \gamma_1) \right\|_2^2 + \lambda \|\alpha\|_1,
$$
where $P_{\widehat{W}^\perp} := I - P_{\widehat{W}}$.

\vspace{2 mm}
\noindent The first term can be minimized to zero by setting

$$
\widehat{\gamma} = \frac{\widehat{W}^\intercal (Y - M \widehat{\gamma}_1)}{\|\widehat{W}\|_2^2}
$$
because $P_{\widehat{W}} P_M = P_{\widehat{W}}$ and $\widehat{W}^\intercal P_{\widehat{W}} = \widehat{W}^\intercal$. Hence, the problem reduces to

$$
(\widehat{\alpha}, \widehat{\beta}) = \arg\min_{\alpha, \beta} \frac{1}{2} \left\| P_{\widehat{W}^\perp} (P_M Y - M \gamma_1) \right\|_2^2 + \lambda \|\alpha\|_1.
$$
Define

$$
\widetilde{D} := P_{\widehat{W}^\perp} D, \quad \widetilde{Z} := P_{\widehat{W}^\perp} Z,
$$
so that

$$
P_{\widehat{W}^\perp} (P_M Y - M \gamma_1) = P_{\widehat{W}^\perp} P_M Y - \widetilde{Z} \alpha - \widetilde{D} \beta.
$$
For fixed $\alpha$, minimize over $\beta$:

$$
\widehat{\beta} = \frac{\widetilde{D}^\intercal (P_M Y - Z \widehat{\alpha})}{\|\widetilde{D}\|_2^2}.
$$
Substituting $\widehat{\beta}$ back, the problem for $\alpha$ becomes

$$
\widehat{\alpha} = \arg\min_{\alpha} \frac{1}{2} \left\| P_{\widetilde{D}^\perp} P_{\widehat{W}^\perp} P_M Y - P_{\widetilde{D}^\perp} P_{\widehat{W}^\perp} Z \alpha \right\|_2^2 + \lambda \|\alpha\|_1,
$$
where $P_{\widetilde{D}^\perp} := I - P_{\widetilde{D}}$.

\vspace{2 mm}
Thus, the estimation proceeds via the two-step algorithm:

$$
\begin{aligned}
\widehat{\alpha} &= \arg\min_{\alpha} \frac{1}{2} \left\| Y - P_{\widetilde{D}^\perp} \widetilde{Z} \alpha \right\|_2^2 + \lambda \|\alpha\|_1, \\
\widehat{\beta} &= \frac{\widetilde{D}^\intercal (Y - \widetilde{Z} \widehat{\alpha})}{\|\widetilde{D}\|_2^2}.
\end{aligned}
$$
This completes the proof.

\subsection{Proof of Lemma \ref{lem: median est}}

Under the stated assumptions, we have the following probability limits for the estimators:
$$
\begin{aligned}
\operatorname{plim}(\widehat{\boldsymbol{\Gamma}}_{-(p_z + 1)}) &= \boldsymbol{\alpha} + \boldsymbol{\delta}^*_{-(p_z + 1)} \gamma, \\
\operatorname{plim}(\widehat{\boldsymbol{\delta}}) &= \boldsymbol{\delta}^*.
\end{aligned}
$$
It then follows that for each instrument $j = 1, \ldots, p_z$, the probability limit of the ratio estimator $\widehat{\pi}_j := \widehat{\Gamma}_j / \widehat{\delta}_j$ is
$$
\operatorname{plim}(\widehat{\pi}_j) = \frac{\alpha_j + \delta^*_j \gamma}{\delta^*_j} = \gamma + \frac{\alpha_j}{\delta^*_j}.
$$
Since at most $s_z < p_z/2$ of the $\alpha_j$ are nonzero, it follows that more than 50\% of the $\alpha_j$ equal zero. Hence, for those instruments $j$ with $\alpha_j = 0$, $\operatorname{plim}(\widehat{\pi}_j) = \gamma$. Therefore, more than half of the entries of $\operatorname{plim}(\widehat{\boldsymbol{\pi}})$ are equal to $\gamma$, and thus the median of $\operatorname{plim}(\widehat{\boldsymbol{\pi}})$ equals $\gamma$. By the **continuous mapping theorem**, this implies:
$$
\operatorname{plim}(\widehat{\gamma}^m) = \operatorname{median}(\operatorname{plim}(\widehat{\boldsymbol{\pi}})) = \gamma.
$$
To characterize the limiting distribution of $\widehat{\gamma}^m$, define the vector $\boldsymbol{\zeta}_1 \in \mathbb{R}^{s_z}$ with components $\zeta_j = \alpha_j / \delta^*_j$ for $j = 1, \ldots, s_z$, and let
$$
\boldsymbol{\zeta} = \begin{pmatrix} \boldsymbol{\zeta}_1 \\ \mathbf{0}_{p_z - s_z} \end{pmatrix} \in \mathbb{R}^{p_z}.
$$
Partition $\widehat{\boldsymbol{\pi}} = (\widehat{\boldsymbol{\pi}}_1^\intercal, \widehat{\boldsymbol{\pi}}_2^\intercal)^\intercal$, corresponding to the nonzero and zero components of $\boldsymbol{\alpha}$, respectively. Under standard regularity conditions, the joint asymptotic distribution of $\widehat{\boldsymbol{\pi}}$ is:
$$
\sqrt{n}\left(\widehat{\boldsymbol{\pi}} - (\gamma \mathbf{1}_{p_z} + \boldsymbol{\zeta})\right) \xrightarrow{d} \mathcal{N}(0, \boldsymbol{\Sigma}_\pi),
$$
where $\mathbf{1}_{p_z}$ denotes a $p_z$-dimensional vector of ones.
Since $\widehat{\gamma}^m = \operatorname{median}(\widehat{\boldsymbol{\pi}})$, we can write:
$$
\begin{aligned}
\sqrt{n}(\widehat{\gamma}^m - \gamma) &= \sqrt{n} \left( \operatorname{median}(\widehat{\boldsymbol{\pi}}) - \gamma \right) \\
&= \operatorname{median} \left( \sqrt{n} (\widehat{\boldsymbol{\pi}} - \gamma \mathbf{1}_{p_z}) \right).
\end{aligned}
$$
Now observe that
$$
\sqrt{n} (\widehat{\boldsymbol{\pi}} - \gamma \mathbf{1}_{p_z}) =
\begin{pmatrix}
\sqrt{n} (\widehat{\boldsymbol{\pi}}_1 - (\gamma \mathbf{1}_{s_z} + \boldsymbol{\zeta}_1)) + \sqrt{n} \boldsymbol{\zeta}_1 \\
\sqrt{n} (\widehat{\boldsymbol{\pi}}_2 - \gamma \mathbf{1}_{p_z - s_z})
\end{pmatrix}.
$$
Since the lower block has at least $p_z - s_z > p_z/2$ elements and converges in distribution to a centered normal distribution (as $\alpha_j = 0$ for those indices), the median is asymptotically driven by this majority block. It then follows that
$$
\sqrt{n}(\widehat{\gamma}^m - \gamma) \xrightarrow{d} q_{[l], p_z - s_z},
$$
where $q_{[l], p_z - s_z}$ denotes the limiting distribution of the median of a Gaussian vector with mean zero and covariance derived from the subvector $\widehat{\boldsymbol{\pi}}_2$.

\subsection{Proof of Proposition \ref{prop: limiting distribution}}
Recall that
$$
\widehat{\beta}_{or} = \frac{\mathbf{D}^{\intercal}\mathbf{P_{\widehat{N}_A^{\perp}}}\mathbf{Y}}{\mathbf{D}^{\intercal}\mathbf{P_{\widehat{N}_A^{\perp}}}\mathbf{D}}
$$
where $\mathbf{N}_A := (\mathbf{Z}_A \quad \mathbf{W})$ and $\widehat{\mathbf{N}}_A := \mathbf{P_M} \mathbf{N}_A$ is the projection of the invalid TCPs and $W$ onto the columnspace of $\mathbf{M}$. It is well known that under standard assumptions,
\begin{equation}
    \sqrt{n}(\widehat{\beta}_{or} - \beta) \to_d N(0, \sigma^2_{or})
\end{equation}
Now we define the post-adaptive LASSO estimator as
\begin{equation}
\widehat{\beta}_{post} := \frac{\mathbf{D}^\intercal \mathbf{P_{\widehat{N}_{ad}^\perp}} \mathbf{Y}}{\mathbf{D}^\intercal \mathbf{P_{\widehat{N}_{ad}^\perp}} \mathbf{D}},
\end{equation}
where $\widehat{\mathbf{N}}_{ad} := \left(\mathbf{Z}_{\widehat{A}_{ad}} \quad \widehat{\mathbf{W}}\right)$ with $\widehat{A}_{ad}$ being the set of invalid TCPs selected by the adaptive LASSO procedure. We can now decompose the post-adaptive LASSO estimator as follows :
$$
\sqrt{n}\left(\widehat{\beta}_{post} - \beta\right) = \sqrt{n}\left(\widehat{\beta}_{or} - \beta\right)\mathbf{1}\{\widehat{A}_{ad} = A\} + \sum_{A^{\prime} \neq A}\sqrt{n}\left(\widehat{\beta}_{post} - \beta\right)\mathbf{1}\{\widehat{A}_{ad} = A^{\prime}\}
$$
Now by selection consistency property of adaptive LASSO we have $plim\left(\mathbf{1}\{\widehat{A}_{ad} = A\}\right) = 1$ and $plim\left(\mathbf{1}\{\widehat{A}_{ad} = A^{\prime}\}\right) = 1$ if $A^{\prime} \neq A$. By Slutsky's theorem, 
$$
\sqrt{n}\left(\widehat{\beta}_{or} - \beta\right)\mathbf{1}\{\widehat{A}_{ad} = A\} \to_d N(0, \sigma^2_{or}).
$$
Further, $plim\left(\sqrt{n}\left(\widehat{\beta}_{post} - \beta\right) - \sqrt{n}\left(\widehat{\beta}_{or} - \beta\right)\mathbf{1}\{\widehat{A}_k{ad} = A\}\right) = 0$. Therefore,
$$
\sqrt{n}\left(\widehat{\beta}_{post} - \beta\right) \to_d N(0, \sigma^2_{or})
$$

\subsection{Proof of Theorem \ref{thm: lim dist median est}}
Without loss of generality, assume that among the $p_w$ candidate OCPs, the first $s_w$ are invalid. For each $k = 1, \ldots, p_w$, when $W_k$ is a valid OCP, Proposition \ref{prop: limiting distribution} ensures that

$$
\sqrt{n}\left(\widehat{\beta}^k_{post} - \beta\right) \xrightarrow{d} \mathcal{N}\left(0, \sigma_{or}^2\right).
$$
We now focus on the case where $W_k$ is invalid. Starting from the underlying model,

$$
\mathbb{E}(Y \mid D, \mathbf{Z}, U) = \beta D + \boldsymbol{\alpha}^{\intercal} \mathbf{Z} + \beta_U U,
$$
which can be expressed for observations $i = 1, \ldots, n$ as

$$
Y_i = D_i \beta + \mathbf{Z}_{i\cdot}^{\intercal} \boldsymbol{\alpha} + U_i \beta_U + \epsilon_{Y_i},
$$
with $\mathbb{E}(\epsilon_Y \mid D, \mathbf{Z}, U) = 0$.

\vspace{2 mm}
The estimator $\widehat{\beta}^k_{post}$ is given by

$$
\widehat{\beta}^k_{post} = \frac{\mathbf{D}^\intercal \mathbf{P}_{\widehat{\mathbf{N}}_{ad}^{k \perp}} \mathbf{Y}}{\mathbf{D}^\intercal \mathbf{P}_{\widehat{\mathbf{N}}_{ad}^{k \perp}} \mathbf{D}},
$$
where $\widehat{\mathbf{N}}_{ad}^k := \left( \mathbf{Z}_{\widehat{A}_{ad}^k} \quad \widehat{\mathbf{W}}_k \right)$ and $\widehat{A}_{ad}^k$ is the set of invalid TCPs identified by adaptive LASSO treating $W_k$ as valid.

\vspace{2 mm}
Expanding, we have

$$
\widehat{\beta}^k_{post} = \beta + \frac{\mathbf{D}^\intercal \mathbf{P}_{\widehat{\mathbf{N}}_{ad}^{k \perp}} \mathbf{Z}_A}{\mathbf{D}^\intercal \mathbf{P}_{\widehat{\mathbf{N}}_{ad}^{k \perp}} \mathbf{D}} \boldsymbol{\alpha}_A + \frac{\mathbf{D}^\intercal \mathbf{P}_{\widehat{\mathbf{N}}_{ad}^{k \perp}} U}{\mathbf{D}^\intercal \mathbf{P}_{\widehat{\mathbf{N}}_{ad}^{k \perp}} \mathbf{D}} \beta_U + \frac{\mathbf{D}^\intercal \mathbf{P}_{\widehat{\mathbf{N}}_{ad}^{k \perp}} \epsilon_Y}{\mathbf{D}^\intercal \mathbf{P}_{\widehat{\mathbf{N}}_{ad}^{k \perp}} \mathbf{D}}.
$$
By the Law of Large Numbers, there exist constants $\{c_{1,k}\}_{1}^{s_w}$ and vectors of constants, $\{\mathbf{c}_{2,k}\}_{1}^{s_w}$ such that

$$
\frac{\mathbf{D}^\intercal \mathbf{P}_{\widehat{\mathbf{N}}_{ad}^{k \perp}} U}{\mathbf{D}^\intercal \mathbf{P}_{\widehat{\mathbf{N}}_{ad}^{k \perp}} \mathbf{D}} \to c_{1,k} \quad \text{and} \quad \frac{\mathbf{D}^\intercal \mathbf{P}_{\widehat{\mathbf{N}}_{ad}^{k \perp}} \mathbf{Z}_A}{\mathbf{D}^\intercal \mathbf{P}_{\widehat{\mathbf{N}}_{ad}^{k \perp}} \mathbf{D}} \to \mathbf{c}_{2,k}^{\intercal}.
$$
Since $W_k$ is invalid, it fails to account for the confounding effect of $U$, ensuring $c_{1,k} \neq 0$. Moreover, the lack of selection consistency for $\widehat{\boldsymbol{\alpha}}^k_{ad}$ implies $plim\left( \mathbf{1}\{\widehat{A}^k_{ad} = A\} \right) \neq 1$, so $c_{2,k} \neq 0$.

\vspace{2 mm}
\noindent Applying the Central Limit Theorem, we find

$$
\sqrt{n} \cdot \frac{\mathbf{D}^\intercal \mathbf{P}_{\widehat{\mathbf{N}}_k^\perp} \epsilon_Y}{\mathbf{D}^\intercal \mathbf{P}_{\widehat{\mathbf{N}}_k^\perp} \mathbf{D}} \xrightarrow{d} \mathcal{N}(0, \sigma_k^2),
$$
where $\sigma_k^2 \neq \sigma_{or}^2$ due to $\widehat{A}_{ad,k} \not\to A$.

\vspace{2 mm}
\noindent Combining these results yield

$$
\sqrt{n} \left( \widehat{\beta}^k_{post} - \left( \beta + c_{1,k} \beta_U + \mathbf{c}_{2,k}^{\intercal} \boldsymbol{\alpha}_A\right) \right) \xrightarrow{d} \mathcal{N}(0, \sigma_k^2).
$$
Define the vector of post-adaptive LASSO estimators for all candidate OCPs as

$$
\widehat{\boldsymbol{\beta}}^{(p_w)}_{post} := \left( \widehat{\beta}^1_{post}, \ldots, \widehat{\beta}^{p_w}_{post} \right)^\intercal.
$$
Then,

$$
\sqrt{n} \left( \widehat{\boldsymbol{\beta}}^{(p_w)}_{post} - \left( \beta \mathbf{1}_{p_w} + \boldsymbol{\zeta} \right) \right) \xrightarrow{d} \mathcal{N}(0, \Sigma_{p_w}),
$$
where

$$
\boldsymbol{\zeta} = \left( \boldsymbol{\zeta}_1, \mathbf{0}_{p_w - s_w} \right), \quad \text{with} \quad \boldsymbol{\zeta}_1 = \left( c_{1,k} \beta_U + \mathbf{c}_{2,k}^{\intercal} \boldsymbol{\alpha}_A \right)_{k=1}^{s_w}.
$$
Hence, if more than half of the candidate OCPs are valid, the median of $\widehat{\boldsymbol{\beta}}^{(p_w)}_{post}$ provides a valid estimator, satisfying

$$
\sqrt{n} \left( \operatorname{median} \left\{ \widehat{\boldsymbol{\beta}}^{(p_w)}_{post} \right\} - \beta \right) \xrightarrow{d} q_{[l], p_w - s_w},
$$
where $q_{[l], p_w - s_w}$ denotes the $l$-th order statistic of the limiting normal distribution of

$$
\sqrt{n} \left( \left\{ \widehat{\beta}^k_{post} \right\}_{k=s_w+1}^{p_w} - \beta\mathbf{1}_{p_w - s_w} \right).
$$

\printbibliography

\end{document}